\documentstyle[preprint,prc,aps,floats,epsf]{revtex}

\def\thalf{{\textstyle{\frac{1}{2}}}}
\def\thrhalf{{\textstyle{\frac{3}{2}}}}
\def\toneth{{\textstyle{\frac{1}{3}}}}
\def\ttwoth{{\textstyle{\frac{2}{3}}}}
\def\tfourth{{\textstyle{\frac{4}{3}}}}
\def\tquar{{\textstyle{\frac{1}{4}}}}
\def\ton{{\textstyle{\frac{1}{9}}}}
\def\ttfeo{{\textstyle{\frac{25}{81}}}}
\def\mo{Moj\v{z}i\v{s}}
\def\pin{$\pi N$}
\def\ch{ChPT}
\def\hch{HBChPT}
\newcommand{\ord}[1]{${\cal O}(Q^{#1})$}
\begin{document}

%
\draft

\preprint{\vbox{\hfill NUC-MINN-97/8-T}}

\title{Pion--Nucleon Scattering in a New Approach to 
       Chiral Perturbation Theory}

\author{Paul J. Ellis and Hua-Bin Tang}
\address{School of Physics and Astronomy,
     University of Minnesota, Minneapolis, MN\ 55455.}

\date{September 15, 1996}
\maketitle

\begin{abstract}
We study pion--nucleon scattering  with a chiral lagrangian
of pions, nucleons, and $\Delta$-isobars.
The scattering amplitude is evaluated to one-loop $Q^3$ order,
where $Q$ is a generic small momentum, using a new approach which is 
equivalent to heavy baryon chiral perturbation theory.
We obtain a good fit to the experimental phase shifts 
for pion center-of-mass kinetic energies up to 100 MeV.
A sigma term greater than 45 MeV is favored, but the value
is not well determined.
\end{abstract}

\vspace{20pt}


\section{Introduction}

The pion and the nucleon play a central role in low-energy physics and 
there is a wealth of scattering data which allow us to test the application of
effective Lagrangians. Several relativistic phenomenological  models 
\cite{OLSSON75,BOFINGER91,PEARCE91,GROSS93,GOUDSMIT94,ELLIS96}
exist, which provide reasonably good fits to the experimental \pin\ phase 
shifts. In these models either the Bethe-Salpeter equation is solved
approximately or the $K$-matrix method is used to unitarize the tree
amplitudes. Such models, however, do not offer a systematic approximation 
scheme.

Chiral perturbation theory (\ch) \cite{WEINBERG79,GASSER84} is a more 
attractive 
approach because  it not only embodies chiral symmetry, which is 
fundamental to low-energy physics, but it also 
offers a systematic expansion in powers of the
momentum. Further it  ensures unitarity order by order. Gasser and Leutwyler
\cite{GASSER84} have shown that ChPT works
nicely for mesons; however, the power counting fails when baryons are
introduced \cite{GASSER88}. The power counting can be restored in 
heavy baryon chiral perturbation theory
(\hch)\cite{JENKINS91} where the heavy components of the  baryon fields 
are integrated out.  Here we shall adopt a 
different approach\cite{TANG96} which effectively removes the heavy fields 
after constructing the Feynman diagrams. This new approach  preserves 
the power counting
and gives results in agreement with \hch, at least to the order considered.

Peccei \cite{PECCEI63} was the first to use a chiral lagrangian 
to calculate the \pin\ scattering lengths 
near threshold. Calculations with 
the more modern approach of \hch\ 
are discussed by Bernard {\it et al.} in a review \cite{BERNARD95} and 
in Refs. \cite{BERNARD95B,BERNARD96}. Particularly interesting 
is the recent calculation  by \mo\ \cite{MO97} of the full amplitude  
to ${\cal O}(Q^3)$, with $Q$ a generic small momentum scale.
\mo\ employs just pion and nucleon fields.
Since there are a number of unknown  parameters, scattering 
lengths and effective ranges alone are not a stringent test of the 
approach. We mention that Datta and Pakvasa \cite{DATTA96} have recently
used the results of \mo\ to discuss low-energy scattering. However a fit
to the phase shifts out to center-of-mass (c.m.) energies 
in the $\Delta$ resonance region is desirable. This requires explicit 
consideration of the $\Delta$ field itself, along with the pion and nucleon 
fields. Here we shall carry out this program by evaluating all diagrams up to
one-loop $Q^3$ order.

The organization of this paper is as follows. In Sec. II we write down our 
effective lagrangian and establish our notation. In Sec. III we describe
our method of separating out the soft contributions, which need to 
be calculated, from the hard contributions, which are subsumed in the 
parameters. Fermion loops and the implications for power counting are 
also discussed here.
The formalism for \pin\ scattering is given in Sec. IV, with a listing of 
the expressions for the loop diagrams relegated to the Appendix. The 
calculated phase shifts are compared with the data in Sec. V, where we also 
discuss the effective $\pi NN$ and $\pi N\Delta$ couplings as well as
the pion--nucleon sigma term. Our conclusions are given in Sec. VI.

\section{Effective Lagrangian}

Before writing down an effective chiral lagrangian we need to 
define our notation, much of which is fairly standard.
The Goldstone pion fields $\pi^a(x)$,
with $a=1,\, 2,\, {\rm and}\, 3\,$,
form an isotriplet that can be written in terms of an $SU(2)$ matrix:
\begin{equation}
 U(x) \equiv \xi ^2 =   \exp (2i\pi(x)/ f_{\pi})
                              \ , \label{eq:U-def}
\end{equation}
where $f_{\pi} \approx 93\,\mbox{MeV}$ 
is the pion decay constant and the pion field is compactly
written as 
$\pi(x) \equiv \bbox{\pi}(x)\, \bbox{\cdot}\,{1\over 2}\bbox{\tau}$,
with $\tau^a$  being the Pauli matrices.
The matrix $U$ is the standard exponential representation and
the ``square root'' representation in terms of $\xi$  is
particularly convenient for including heavy fields in the chiral
lagrangian.
The isodoublet nucleon field is represented by a column matrix
\begin{equation}
N=\left (\,
     \begin{array}{c} 
         p \\ 
         n
      \end{array} \right )\ ,
\end{equation}
where  $p$ and $n$ are the proton and neutron fields
respectively. 
The $\Delta$ is a spin-$\thrhalf$ and isospin-$\thrhalf$ particle
represented by an isoquadruplet field:
\begin{equation}
\Delta_\mu = \left( 
                    \begin{array}{l}
                       \Delta_\mu^{++}\\
                       \Delta_\mu^{+}\\
                       \Delta_\mu^{0}\\
                       \Delta_\mu^{-}
                    \end{array}
             \right)    \ .   \label{eq:Del}
\end{equation}
It is convenient to introduce  an isovector field
$\bbox{\Delta}_\mu = \bbox{T} \Delta_\mu$
in terms of the standard $2\times 4$
isospin $\thrhalf$ to $\thalf$  transition matrix:
\begin{equation}
\langle {\case1/2}\, t |
   \bbox{T} |{\case3/2}\, t_\Delta \rangle
  \equiv \sum_\lambda \langle 1\, \lambda \,{\case1/2} \,t |
    {\case3/2} \,t_\Delta\rangle \bbox {e}_\lambda
                     \ , \label{eq:T}
\end{equation}
where the isospin spherical unit vectors are $\bbox{e}_0 = \bbox{e}_z$ and
$\bbox{e}_{\pm 1}=\mp (\bbox{e}_x \pm i \bbox{e}_y)/\sqrt{2}$.
Explicitly, the components are
\begin{eqnarray}
     \Delta_\mu^1 + i  \Delta_\mu^2 & = &
             {1\over \sqrt{3}} \left( \!\!
                    \begin{array}{c}
                      \sqrt{2} \Delta_\mu^{0}\\
                      \sqrt{6} \Delta_\mu^{-}
                    \end{array}  \!\!
             \right)
             \ ,        \\[3pt]
    \Delta_\mu^1 - i  \Delta_\mu^2 & = &
            - {1\over \sqrt{3}} \left( \!\!
                    \begin{array}{c}
                      \sqrt{6} \Delta_\mu^{++}\\
                      \sqrt{2} \Delta_\mu^{+}
                    \end{array}  \!\!
             \right)
                \ ,     \\[3pt]
     \Delta_\mu^3 &=& \sqrt{2\over 3} \left( \!\!
                    \begin{array}{c}
                      \Delta_\mu^{+}\\
                      \Delta_\mu^{0}
                    \end{array}\!\!
             \right)       \ .
\end{eqnarray}

   Following Callan {\it et al.} \cite{CCWZ69} we define a  nonlinear 
realization of the chiral group $SU(2)_{\rm L}\otimes SU(2)_{\rm R}$ 
such that, for arbitrary global matrices $L \in SU(2)_{\rm L}$ 
and $R \in SU(2)_{\rm R}$, we have the mapping
\begin{equation}
L\otimes R:\ \ \ (\xi, N, \Delta_\mu)\longrightarrow 
        (\xi', N', \Delta'_\mu)   
          \ ,     \label{eq:nonlr}
\end{equation}
where
\begin{eqnarray}
\xi'(x) &=& L \xi(x) h^{\dagger}(x) = h(x) \xi(x) R^{\dagger}
               \ , \label{eq:Xitrans} \\[4pt]
 N'(x) &=& h(x)N(x)  \ , \label{eq:Ntrans} \\[4pt]
 \bbox{\Delta}'_\mu(x) &=& \thalf 
              h \, {\rm tr}(h\bbox{\tau}
            h^\dagger \bbox{\tau})\bbox{\cdot} \bbox{\Delta}_\mu (x)
              \ .       \label{eq:Dtrans}
\end{eqnarray}
As usual, the matrix $U$ transforms as $U'(x) = LU(x)R^{\dagger}$.
The second equality in Eq.~(\ref{eq:Xitrans}) defines
$h(x)$ implicitly as a function of $L$, $R$, and the local pion fields:
$h(x)=h(\bbox{\pi}(x),L,R)$.
The pseudoscalar nature of the pion field 
implies  $h(x)\in SU(2)_{\rm V}$, with
$ SU(2)_{\rm V}$ the unbroken vector  subgroup
of $SU(2)_{\rm L} \otimes SU(2)_{\rm R}$.
The nucleon  transforms linearly under
$SU(2)_{\rm V}$ as an isodoublet. While the isodoublet
components of the isovector $\bbox{\Delta}_\mu$ transform linearly
in the same way as the  nucleon field, the isovector itself is
further rotated by the $O(3)$  transformation 
$\thalf {\rm tr}(h\bbox{\tau}h^\dagger \bbox{\tau})$.

Interaction terms invariant under the nonlinear chiral transformation
may be conveniently constructed in terms of an axial vector field
$a_{\mu}(x)$ and a polar vector field $v_{\mu}(x)$ defined as
\begin{eqnarray}
a_{\mu}     & \equiv & -{i \over 2}(\xi^{\dagger} \partial_{\mu} \xi -
    \xi \partial_{\mu}\xi^{\dagger} )
          = a_{\mu}^\dagger= \thalf\bbox{a}_\mu\bbox{\cdot\tau}
     ={1\over f_\pi}\partial_{\mu}\pi
             - {1\over 3f_\pi^3}  \pi [\pi,\partial_{\mu}\pi] +\cdots    
          \ ,\label{eq:adef}  \\[4pt]
v_{\mu}     & \equiv & 
           -{i \over 2}(\xi^{\dagger} \partial_{\mu} \xi +
    \xi \partial_{\mu}\xi^{\dagger} )
         = v_{\mu} ^\dagger   = \thalf\bbox{v}_\mu\bbox{\cdot\tau}
         =-{i\over 2f_\pi^2}
           \left(1-{\pi^2\over 3f_\pi^2}\right) [\pi,\partial_{\mu}\pi]+ \cdots 
            \ , \label{eq:vdef}
\end{eqnarray}
both of which contain one derivative. 
The polar vector field transforms inhomogeneously and the axial
vector field transforms homogeneously:
\begin{eqnarray}
v_\mu'   &= &   h v_\mu h^{\dagger}
                   -ih\partial_\mu h^{\dagger}
               \ , \\[3pt]
a_\mu'   &= &    h a_\mu h^{\dagger} \ . \label{eq:homog}
\end{eqnarray}

To maintain chiral invariance, instead of an ordinary
derivative $\partial^\mu$, one uses a covariant derivative
${\cal D}_\mu$ on the nucleon and $\Delta$ fields.
These are defined by
\begin{eqnarray}
{\cal D}_\mu N &=& \partial_\mu N + i v_\mu N \ , \\
{\cal D}_\mu \bbox{\Delta}_\nu 
            &=& \partial_\mu\bbox{\Delta}_\nu
                      + i  v_\mu  \bbox{\Delta}_\nu
                      -   \bbox{v}_\mu \times \bbox{\Delta}_\nu\ .
\end{eqnarray}
We also use the following definitions involving two and three derivatives
on the pion field
\begin{eqnarray}
v_{\mu\nu} &=& \partial_{\mu} v_{\nu} -\partial_{\nu}
      v_{\mu} + i [v_{\mu}, v_{\nu}] 
       = - i [a_\mu, a_\nu]   \ , \\[4pt]
    D_\mu a_{\nu} &=& \partial_\mu a_{\nu}
                            + i [v_\mu, a_{\nu}]  \ , \\[4pt]
    D_\sigma v_{\mu\nu} &=& \partial_\sigma v_{\mu\nu}
                            + i [v_\sigma, v_{\mu\nu}] \ ,
\end{eqnarray}
all of which transform homogeneously in the same way 
as  $a_{\mu}$ in Eq.~(\ref{eq:homog}).

To write a general effective lagrangian, we need an organizational scheme
for the interaction terms. We organize the lagrangian in increasing powers
of the fields and their derivatives. Specifically, as in 
Refs.~\cite{ELLIS96,FST97},                           
we assign to each interaction term a size of order $Q^{\alpha}$ with
\begin{equation}
\alpha = d+{n\over 2}  \ ,   \label{eq:order}
\end{equation}
where $d$  is the number  of derivatives on the pion field or pion 
mass ($m_\pi$) factors,  and $n$ is the number of fermion fields 
in the interaction term. 
That $\alpha$ is a characteristic of the interaction term
is suggested by 
Weinberg's power counting \cite{WEINBERG90}, which we discuss
in Sec.~III below.
Derivatives on the nucleon fields are not counted in $d$ because they
will generally be associated with the nucleon mass and not with the small
momentum $Q$. In fact, as Krause \cite{KRAUSE90} has argued,
it is $i\rlap/{\mkern-3mu {\cal D}} -M$
that is of \ord{1}. Krause also counts  $\gamma_\mu\gamma_5$ to be of 
\ord{0} and a single factor of $\gamma_5$ to be of \ord{1}.
We adopt this counting for organizing
the lagrangian, although we have argued \cite{ELLIS96} that counting
a  single $\gamma_5$ factor to be of  \ord{1} is not
precise. Our scheme allows a uniform organization of the pion 
self-interaction terms and those involving the heavy fermions.
It differs from the ``standard labelling" of Gasser, Sainio 
and \v{S}varc \cite{GASSER88}
where the number of fermion fields is not included.

Taking into account
chiral symmetry, Lorentz invariance, and parity conservation, we
may write the lagrangian through quartic order ($\alpha \leq 4$) as
the sum of the order $Q^2$, $Q^3$, and $Q^4$ parts:
\begin{equation}
{\cal L} = {\cal L}_2 + {\cal L}_3 + {\cal L}_4 + \Delta{\cal L} \ ,
\end{equation}
where $\Delta{\cal L}$ represents the counterterms, which can also
be organized in powers of $Q$. 
We will adopt the counterterm method of renormalization, so
we start with the 
physical masses and couplings and add the necessary counterterms.
For simplicity, in this first investigation we will not explicitly 
identify the finite and divergent pieces of the various counterterms.
The order $Q^2$ part of the lagrangian is
\begin{eqnarray}
  {\cal L}_2 &=& \overline{N} ( i\rlap/{\mkern-3mu {\cal D}}
               +g_{\rm A}\gamma^\mu \gamma_5 a_\mu - M ) N 
           +\tquar f_\pi^2  {\rm tr}\, (\partial_\mu U^\dagger
                         \partial^\mu U)
           +\tquar m_\pi^2f_\pi^2 {\rm tr}\,(U + U^\dagger -2)
       \nonumber  \\
      & & 
          +  \overline{\Delta}_\mu^a
           \Lambda^{\mu\nu}_{ab} \Delta_\nu^b
       + h_{\rm A} \Big ( \overline{\bbox{\Delta}}_\mu
         \bbox{\cdot a}^\mu  N
       + \overline{N}
           \bbox{a}^\mu\bbox{\cdot \Delta}_\mu \Big)
         + \tilde{h}_{\rm A}
             \overline{\Delta}_\mu^{\, a}
                  \gamma^\nu \gamma_5 a_\nu
                    \Delta^\mu_a   \ ,   \label{eq:L2}
\end{eqnarray}
where the isospin indices $a,\, b =  1,\, 2,\, {\rm and}\, 3$,
the trace is taken over the isospin matrices, and
the kernel tensor in the   $\Delta$ kinetic-energy term is
\begin{equation}
    \Lambda^{\mu\nu} = -(i \rlap/{\mkern-3mu {\cal D}} 
                - M_\Delta)g^{\mu\nu} 
              +i(\gamma^\mu {\cal D}^\nu +\gamma^\nu {\cal D}^\mu)
              -\gamma^\mu(i\rlap/{\mkern-3mu {\cal D}}+M_\Delta)\gamma^\nu
                        \ ,
\end{equation}
suppressing isospin indices.                                
Here we have chosen the standard parameter $A=-1$, because
it can be modified by redefinition of the $\Delta$ field with no
physical consequences \cite{NATH70}. In the $\pi N\Delta$ interaction
of Eq.~(\ref{eq:L2}) we have chosen the standard
off-shell $Z$ parameter to have the convenient value of $-\thalf$.
The value of $Z$ has no physical significance since
modifications can be absorbed in the other parameters in the 
lagrangian \cite{TE96}. Similarly
we simplify the $\pi\Delta\Delta$ interaction
by choosing the offshell parameters defined in Ref.~\cite{TE96}
to be $Z_2=-\thalf$ and $Z_3=0$.

With the notation
\begin{equation}
  \stackrel{\leftrightarrow}{\cal D}_\mu
      = {\cal D}_\mu -
         (\stackrel{\leftarrow}{\partial}_\mu - iv_{\mu}) \ ,
\end{equation}
we may write the order $Q^3$ and $Q^4$ parts of ${\cal L}$ as follows:
\begin{eqnarray}
  {\cal L}_3 &=&  { \beta_\pi \over M} \overline{N} N 
         {\rm tr}\, (\partial_\mu U^\dagger \partial^\mu U)
          -{\kappa_\pi\over M} \overline{N}
               v_{\mu\nu}\sigma^{\mu\nu} N
                     \nonumber  \\
      & & +{\kappa_1\over 2 M^2} i\overline{N}
               \gamma_\mu 
               \stackrel{\leftrightarrow}{\cal D}_\nu N
             {\rm tr}\, (a^\mu a^\nu)
          +{\kappa_2\over M} m_\pi^2 \overline{N} N\, 
            {\rm tr}\,(U + U^\dagger -2) +\cdots
                   \ ,     \label{eq:L3}  \\[4pt]
  {\cal L}_4 &=& 
             {\lambda_1\over M} m_\pi^2 \overline{N} \gamma_5
                (U - U^\dagger ) N
             + {\lambda_2\over M^2} \overline{N}
                        \gamma^\mu D^\nu v_{\mu\nu} N
                     \nonumber  \\
      & & 
           +{\lambda_3\over M^2} m_\pi^2 \overline{N} \gamma_\mu
               [a^\mu, \, U - U^\dagger] N
          +{\lambda_4\over 2 M^3} i\overline{N}
               \sigma_{\rho \mu} 
               \stackrel{\leftrightarrow}{\cal D}_\nu N
             {\rm tr}\, (a^\rho D^\mu a^\nu)
                     \nonumber  \\
      & &
          +{\lambda_5\over 16M^4} i\overline{N}
         \gamma_\rho 
            \{\stackrel{\leftrightarrow}{\cal D}_\mu,
            \stackrel{\leftrightarrow}{\cal D}_\nu\} \tau^a N 
          \,{\rm tr}\,(\tau^a[D^{\rho}a^\mu,a^\nu])
              \nonumber  \\
      & &
         +{\lambda_6\over M^2} m_\pi^2
          \Big \{ \overline{\bbox{\Delta}}_\mu
         \bbox{\cdot} {\rm tr}[i\partial^\mu  (U - U^\dagger )
            \bbox{\tau} ]  N
             + \overline{N} {\rm tr}[i\partial^\mu  (U - U^\dagger )
            \bbox{\tau} ] \bbox{\cdot \Delta}_\mu \Big\}
            +\cdots
                   \ .     \label{eq:L4}
\end{eqnarray}
In Eqs.~(\ref{eq:L3}) and (\ref{eq:L4}) the ellipsis represents terms 
that do not contribute to the $\pi N$ scattering amplitude. For example,
the ellipsis in ${\cal L}_4$ includes the usual pion self-interaction
terms with four derivatives. The $m_\pi^2$ factors in $ {\cal L}_4$ are 
introduced to correctly count the order of the symmetry breaking terms. 
However, it appears that counterterms of the above form with 
$m_\pi^2$ replaced by $\delta^2$ are needed even in the chiral limit 
($m_\pi=0$). Here $\delta=M_\Delta-M$ is
the $\Delta$-nucleon mass difference. We have 
applied naive dimensional analysis \cite{FST97,GEORGI93}
to the terms in Eqs.~(\ref{eq:L3}) and (\ref{eq:L4}) so as to expose
the dimensional factors. As a result, we expect the parameters 
to be of order unity.

Using the pion and nucleon equations of motion 
\cite{WEINBERG90,GEORGI91,ARZT95}, 
we have also simplified 
the contact terms listed in Ref.~\cite{GASSER88}.
For example, we reduce the \ord{3} term 
$ \overline{N}   \stackrel{\leftrightarrow}{\cal D}_\mu 
               \stackrel{\leftrightarrow}{\cal D}_\nu\! N
             \, {\rm tr}\, (a^\mu a^\nu)
$
to the sum of the \ord{3} $\kappa_1$ term, 
the \ord{4} $\lambda_4$ term, and higher-order terms which we omit.
As a result we have the  minimum number of independent terms 
contributing to the $\pi N$ scattering amplitude up to \ord{3}.
Note that the isoscalar-scalar $\phi$
and isovector-vector $\rho$ fields given in Ref.~\cite{FST97}
have been integrated out. Their effects show up 
in the contact terms $\beta_{\pi},\ \kappa_2$ and 
$\lambda_2$. For example, in terms of the 
$\rho\pi\pi$ coupling ($g_{\rho\pi\pi}$) and the
$\rho N N$ coupling ($g_\rho$), the rho gives a contribution
to the $\lambda_2$ parameter of 
$-2 g_{\rho\pi\pi}g_\rho M^2f_\pi^2/m_\rho^4$.

\section{Chiral Perturbation Theory with Baryons}

\subsection{Hard and Soft Contributions}

Given the effective lagrangian, one can derive the Feynman rules
and carry out perturbative calculations of physical quantities in
the standard way. However, as shown by Gasser, Sainio 
and \v{S}varc \cite{GASSER88}, the loop expansion
no longer corresponds directly to the momentum expansion when 
we have heavy fermions. One way to overcome this difficulty is 
\hch\ \cite{JENKINS91}, 
where the heavy components of the fermion fields are integrated out
so that their effects on physical quantities only show up in the
parameters of the lagrangian.
Alternatively one can construct an explicitly nonrelativistic 
lagrangian\cite{WEINBERG90}.

We propose a different procedure here which involves manipulating the 
Feynman diagrams themselves 
(see also the recent discussion of Gasser \cite{GASSER97}). 
First we obtain the Feynman rules in the standard way.
Then we separate the loop contributions into those from soft
and hard momenta. We keep the soft contributions explicitly. These will 
have both real and imaginary parts in general, the latter being needed to
maintain unitarity order by order. As for the hard contributions,
we implicitly absorb them into the coefficients of the lagrangian. 
As we will see this procedure preserves a systematic power 
counting scheme.

Specifically, we represent the hard momentum scale by the 
nucleon mass $M$. Other quantities of this
order include the $\Delta$ mass $M_\Delta$ and the factor $4\pi f_\pi$
with $4\pi$ coming from a loop integral \cite{MANOHAR84}.
The soft momentum scale is denoted by $Q$, where $Q\ll M$. Quantities 
of this order 
are the pion mass $m_\pi$, the pion momentum, and
the mass splitting $\delta$ between the nucleon and the 
$\Delta$-isobar. Also, as in \hch, we are interested in applications 
where the three-momenta of the external nucleons are of order $Q$.

For the present we consider any loop diagram without fermion loops:
we shall consider diagrams with fermion loops later. We 
obtain the unrenormalized soft part of the diagram
by applying the following rules to the loop integral:
\begin{itemize}
\item[1.] Take the loop momenta of the pion lines to be of order $Q$.
\item[2.] Make a covariant $Q/M$ expansion of the integrand.
\item[3.] Exchange the order of the integration and 
           summation of the power series.
\end{itemize}
Rule 1 ensures that the exchanged pions have soft momenta. As a result
internal baryon lines will be nearly on shell throughout the diagram.
Rule 2 implies that a covariant expansion of the baryon propagators in 
the integrand is made, which maintains the Lorentz invariance of the soft 
part. In Rule 3 the exchange of the order of integration and summation 
changes the result in general. Indeed the purpose of this maneuver is to 
remove the poles in the baryon propagators at hard loop momenta of order $M$. 
Clearly this is achieved because after application of Rule 2 the only 
poles in individual terms of the 
series are located in the soft momentum region of ${\cal{O}}(Q)$. 
Of course the soft part obtained from our rules
still contains ultraviolet divergences 
in the form of poles at $d=4$ in dimensional regularization. We remove
these divergences with the standard method of renormalization. 
Formally, if we denote
the original integral by $\cal{I}$, the unrenormalized soft part 
by $\hat{S}\cal{I}$ and
the renormalization operator by $\hat{R}$, the final
renormalized soft part is $\hat{R}\hat{S}\cal{I}$. 
This is the physical loop contribution.
As discussed in the next section,
the soft loop contributions allow for a systematic power counting.

As for the part of the original integral $\cal{I}$ that is discarded, namely
${\cal{I}}-\hat{R}\hat{S}\cal{I}$, we call it the hard part. This
hard part contains contributions from poles of the integrand at hard momenta.
Thus, we should be able to absorb this part into the coefficients of the
lagrangian. As Lepage\cite{LEPAGE89} has argued from
the uncertainty principle,
large momenta correspond to short distances that are
tiny compared with the wavelengths of the external particles,
so the interactions must be local.

\begin{figure}
 \setlength{\epsfxsize}{3in}
  \centerline{\epsffile{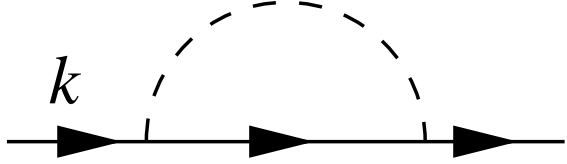}}
\vspace*{.2in}
\caption{A one-loop nucleon self-energy diagram. The dashed line represents
a pion propagator and the solid line a nucleon propagator.}
 \label{fig:one}
\vspace*{.4in}
\end{figure}

At this stage a concrete example is useful.
Thus, we evaluate the 
nucleon self-energy diagram shown in Fig.~\ref{fig:one}.
From standard Feynman rules we obtain the self-energy:
\begin{equation}
\Sigma_{\pi\rm N}(k) = 
             { 3g_{\rm A}^2 \over 4 f_\pi^2 }\, i\mu^{4-d}
         \!   \int\!{{\rm d}^d{\ell}\over ( 2\pi)^d}
           {  \rlap/{\mkern-1mu}\ell \gamma_5
              G(k+\ell)\,
              \rlap/{\mkern-1mu}\ell \gamma_5
              \over \ell^2 - m_\pi^2 + i \epsilon }
                      \ , \label{sigpiN}
\end{equation}
where the free nucleon propagator
\begin{equation}
  G(k) =         { 1 \over \rlap/{\mkern-2mu}k
                       - M + i \epsilon }
             \ , 
\end{equation}
and $\mu$ is the scale of dimensional regularization.
We obtain the soft part of $\Sigma_{\pi\rm N}$
by first making a covariant $Q/M$ expansion of the integrand, while
taking the pion momentum $\ell$ to be soft. Since we assume
$k$ to be nearly on shell, we can 
make a covariant expansion of the nucleon propagator as
\begin{equation}
  G(k+\ell)
        = {1\over
              2k\cdot \ell+ k^2 - M^2 + i\epsilon}
          \bigg[(\rlap/{\mkern-2mu}k+M)   +\rlap/{\mkern-1mu}\ell 
             -{\ell^2(\rlap/{\mkern-2mu}k
               + M ) \over 2k\cdot \ell+ k^2 - M^2 + i\epsilon}
             +\cdots
          \bigg] \ . \label{eq:expN}
\end{equation}
Here the contribution arising from first term in the square brackets is 
of order $Q^{-1}$, while that due to the second and third terms is of 
order $Q^{-1}\times(Q/M)$. Subsequent terms will involve 
higher powers of $Q/M$. The precise way this expansion is carried out has
to be tailored to the case at hand. For example, in $\pi N$ scattering we have 
$k=(p+q)$ where $p^2=M^2$ and $q^2=m_{\pi}^2$. Then we may let
$k\rightarrow p$ and $\ell\rightarrow(\ell+q)$ in Eq.~(\ref{eq:expN}).

Exchanging the order of the summation and integration in 
Eqs.~(\ref{sigpiN}) and (\ref{eq:expN}) then yields the soft part:
\begin{equation}
\hat{S} \Sigma_{\pi \rm N}(k) 
           = {3\over 4}
                  { g_{\rm A}^2 \over f_\pi^2 }\, i\mu^{4-d}\!
           \int\!{{\rm d}^d{\ell}\over ( 2\pi)^d}
           { 2k\cdot \ell \, \rlap/{\mkern-1mu}\ell 
              -(\rlap/{\mkern-2mu} k + M) \ell^2 
           \over (\ell^2 - m_\pi^2 + i \epsilon )
                 (2k\cdot \ell+ k^2 - M^2 + i\epsilon) } +\cdots
               \ . \label{eq:piNexpn}
\end{equation}
Here we explicitly show the leading contribution to the soft part
$\hat{S} \Sigma_{\pi \rm N}$ which is of order 
$Q^3/M^2$, as would be expected from Weinberg's power 
counting. The ellipsis represents higher-order terms. For 
illustrative purposes, it is useful to sum up the series of soft 
contributions. This can be carried out by noting that in the present 
case, after our exchange of the integration and summation, 
an $\ell^2$ in the numerator of any integrand 
can be replaced with $m_\pi^2$ in dimensional regularization.
Then the exact soft part of 
the one-loop self-energy diagram is
\begin{equation}
\hat{S} \Sigma_{\pi \rm N}(k) 
           = {3\over 4}
                  { g_{\rm A}^2 \over f_\pi^2 }\, i\mu^{4-d}\!
           \int\!{{\rm d}^d{\ell}\over ( 2\pi)^d}
           { (2 k\cdot \ell +m_\pi^2)\rlap/{\mkern-1mu}\ell
               -  (\rlap/{\mkern-2mu} k + M)m_\pi^2
           \over (\ell^2 - m_\pi^2 + i \epsilon )
                 (2k\cdot\ell + k^2 - M^2 + m_\pi^2 + i\epsilon) }
               \ . \label{eq:piN}
\end{equation}
Introducing
\begin{equation}
     \omega \equiv
           {1\over 2\sqrt{k^2} }(k^2-M^2 +m_\pi^2) \ ,\label{eq:omega}
\end{equation}
which is of order $Q$, the integral of Eq.~(\ref{eq:piN})
in the case $|\omega|<m_\pi$ can be written
\begin{eqnarray}
\hat{S}\Sigma_{\pi\rm N} (k) &=&
               {3 g_{\rm A}^2\over 2(4\pi f_\pi)^2}
                \bigg ( { m_\pi^2 \over \sqrt{k^2} }
                      (\rlap/{\mkern-2mu}k + M)
                 - {k^2-M^2 \over k^2} \omega
                  \, \rlap/{\mkern-2mu}k \bigg)
           \bigg[ {\omega\over 2} -  \sqrt{m_\pi^2-\omega^2}
              \, \cos^{-1}\bigg({-\omega\over m_\pi} \bigg)\bigg]
             \nonumber \\[3pt]
             & & \ \ \ \
                - {3 g_{\rm A}^2\over 8(4\pi f_\pi)^2}
     \bigg [ {k^2-M^2 \over k^2} (m_\pi^2-2\omega^2)\rlap/{\mkern-2mu}k
                    +{ 2\omega m_\pi^2 \over \sqrt{k^2} }
                      (\rlap/{\mkern-2mu}k + M)\bigg]
               \bigg( 32\pi^2L +\ln {m_\pi^2 \over \mu^2} 
                           \bigg)
                                \ , \label{eq:Nself}
\end{eqnarray}
where 
\begin{equation}
      L\equiv  {1\over 32\pi^2}
            \Big( {2\over d-4}+\gamma -1 -\ln 4\pi \Big) \ , \label{eq:L}
\end{equation}
with  Euler's constant $\gamma = 0.577\cdots$.

In the modified minimal 
subtraction ($\overline{\rm MS}$) renormalization scheme, we obtain the 
renormalized soft part $\hat{R}\hat{S}\Sigma_{\pi\rm N}$ by including 
counterterm contributions (CTCs) to remove the term proportional to $L$
in Eq.~(\ref{eq:Nself}). We can further ensure that the pole of the 
nucleon propagator is at the physical mass with unit residue by
additional on-shell mass and wavefunction counterterm 
subtractions as detailed in Eq.~(\ref {eq:subnself}) below.
These CTCs can clearly be expanded  in an 
infinite power series in $(\rlap/{\mkern-1mu}k- M)/M$.
Thus, the divergences appear to all orders in the $Q/M$ expansion. 
The divergences can be removed by introducing 
counterterms of the form $
{1\over M^{n-1}}\overline{N}
       (i\rlap/{\mkern-1mu} \partial - M)^n N
$
with $n$ an integer. However, we note that 
we may use the nucleon 
equation of motion\cite{WEINBERG90,GEORGI91,ARZT95} to eliminate
the above
counterterms in favor of interaction terms involving multiple pions
and nucleons. 

We may obtain the leading order $Q^3/M^2$ contribution to
Eq.~(\ref{eq:Nself})
by approximating $\omega\simeq(k^2-M^2)/(2M)$ and setting
$\sqrt{k^2}\simeq M$ in the denominators.
It is not possible to expand this leading-order expression further since
the square root and inverse cosine 
functions in the equation involve $\omega$ and $m_\pi$ which 
are of the same order. Thus, we cannot absorb
this soft part into the parameters of the lagrangian.
This result is consistent with the expectation that the parameters should
contain only high-energy contributions.

The hard part of the self energy can also be evaluated directly.
We find
\begin{eqnarray}
(1-\hat{S})\Sigma_{\pi\rm N} (k)
            &=&   -{3\over 4}
                  { g_{\rm A}^2 \over f_\pi^2 }\, i\mu^{4-d}\!
           \int\!{{\rm d}^d{\ell}\over ( 2\pi)^d}
           { (\rlap/{\mkern-1mu}\ell - \rlap/{\mkern-2mu} k)
             (\rlap/{\mkern-1mu}\ell - M)
               (\rlap/{\mkern-1mu}\ell - \rlap/{\mkern-2mu} k)
           \over (\ell^2 - M^2 + i \epsilon )
                 (2k\cdot\ell - k^2 - M^2 + m_\pi^2 + i\epsilon) }
                         \nonumber \\[3pt]
         &=&
               - {3 g_{\rm A}^2\over 4(4\pi  f_\pi)^2} M^2
               \bigg ( M
                    +{k^2+M^2 \over 2k^2}\rlap/{\mkern-2mu}k
                \bigg)
                 \bigg( 32\pi^2 L +
                          \ln {M^2 \over \mu^2} \bigg)
                  \nonumber \\[3pt]
 &    & 
      + {3 g_{\rm A}^2\over 4(4\pi f_\pi)^2} 
           \Big(\sqrt{k^2}-\omega\Big)
               \bigg[
                 {k^2-M^2 \over k^2} \omega
                  \, \rlap/{\mkern-2mu}k
                -{ m_\pi^2 \over \sqrt{k^2} }
                      (\rlap/{\mkern-2mu}k + M)\bigg]
                  \nonumber \\[3pt]
      &    & \ \ \ \  \
               \times  \bigg[ 32\pi^2 L
                       +   \ln {M^2 \over \mu^2} 
             - 1 + \sum_{l=1}^{\infty} {2  \over 2 l -1}\,
             {(\omega^2 - m_\pi^2)^l \over
                      (\sqrt{k^2}-\omega)^{2l} }\, \bigg]
                      \ . \label{eq:hard}
\end{eqnarray}
Notice that
the integral in the first equation of (\ref{eq:hard}) is dominated by
poles at momenta of ${\cal O}(M)$. The final result
can indeed be expanded in powers of 
$(\rlap/{\mkern-1mu} k-M)/M$ and thereby removed by CTCs.
If this were not done the power counting would be spoilt 
since the first term of the result
is of ${\cal O}(M)$ and
the second is of ${\cal O}(Q^2/M)$, which can be contrasted
with ${\cal O}(Q^3/M^2)$ for the soft part.

While our procedure is plausible, a general proof would require 
consideration of diagrams of arbitrary complexity.
Here we restrict ourselves to one-loop order for which 
it is easy to see that our procedure gives the same
result as HBChPT. Indeed, the expansion of the baryon 
propagator in Eq.~(\ref{eq:expN}) 
generates the same effects as integrating 
out the heavy components of the baryon fields in HBChPT. 
For our example of the nucleon self-energy we can make the connection
with HBChPT by introducing in Eq.~(\ref{eq:piNexpn})
the four velocity $v_{\mu}$ such that 
$k_\mu=Mv_\mu+q_\mu$ with $v_{\mu}v^{\mu}=1$.
Projecting onto the light components by inserting the projection operators
$\thalf(1+\rlap/{\mkern-2mu} v)=[\thalf(1+\rlap/{\mkern-2mu} v)]^2$
fore and aft, and noting that 
\begin{equation}
      \thalf(1+\rlap/{\mkern-2mu} v)\:\rlap/{\mkern-1mu}\ell 
       \gamma_5\:\thalf(1+\rlap/{\mkern-2mu} v) = 2\ell\cdot S
        \:\thalf(1+\rlap/{\mkern-2mu} v)\;,
\end{equation}
where $S_{\mu}=\thalf i\gamma_5\sigma_{\mu\nu}v^{\nu}$, we find
\begin{eqnarray}
\thalf(1+\rlap/{\mkern-2mu} v)\:\hat{S}\Sigma_{\pi \rm N}(k) \:
      \thalf(1+\rlap/{\mkern-2mu} v)
           &=&  \thalf(1+\rlap/{\mkern-2mu} v)\:
         { 3g_{\rm A}^2 \over f_\pi^2 }\, i\mu^{4-d}\!
           \int\!{{\rm d}^d{\ell}\over ( 2\pi)^d}\,
           { (\ell\cdot S)^2
           \over \ell^2 - m_\pi^2 + i \epsilon }
                        \nonumber \\[3pt]
      &    & \ \ \ \  \ \ \ \  \ \ \ \  \  \ \ \ \  \  \ \ \ \  \ 
            \times  {1 \over v\cdot q-  v\cdot \ell + i\epsilon } 
                 +\cdots         \ . \label{hb:piN}
\end{eqnarray}
This is a well-known expression  which follows directly 
from the Feynman rules of \hch\ \cite{BERNARD95,MO97}.
Our expression for the soft part, Eq.~(\ref{eq:Nself}), when projected
onto the light components yields in leading order (or equivalently
in the infinite nucleon mass limit) the result of Bernard {\it et al.}
\cite{BERNARD95,BERNARD92} obtained in HBChPT.

\mo\ \cite{MO97} has
recently calculated diagrams for $\pi N $ scattering
using \hch\ with just pions and nucleons  up to one-loop $Q^3$ order.
The results of our calculation, given in the Appendix,
agree with his (modulo differences in the
parameterization of $U$ and the treatment of finite terms arising from
the product of a $(d-4)$ factor with the $1/(d-4)$ pole,
see below). For simplicity our examples have involved the nucleon          
propagator, but we emphasise that the delta propagator appearing in loop   
diagrams (denoted by an open box in Figs.~\ref{fig:nonvtx} and             
\ref{fig:vtx} below) is treated                                            
in exactly the same way. This is necessary to preserve the power counting  
and it gives results which agree with the HBChPT approach of               
Hemmert et al.\cite{HEMMERT97}.                                            

\subsection{Fermion Loops and Power Counting}

\begin{figure}
 \setlength{\epsfxsize}{3in}
  \centerline{\epsffile{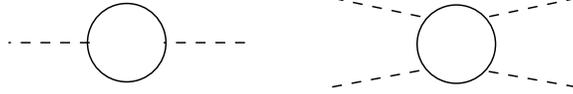}}
\vspace*{.2in}
\caption{Sample diagrams with baryon loops not directly
connected to external baryon lines.}
 \label{fig:two}
\vspace*{.4in}
\end{figure}

Fermion loops were not discussed in the preceding section.
We notice that there are no fermion loops in HBChPT after integrating
out the heavy field components. Here we show that fermion loops
have vanishing soft parts in our approach so that they can be ignored.
First consider fermion loops that are not directly connected 
to external fermion lines, such as those in Fig.~\ref{fig:two}.
We will work with nucleons for simplicity, although similar arguments 
can be given for $\Delta$'s. If we generalize our previous rules by 
taking the loop momentum to be of ${\cal O}(Q)$,
we can expand the propagators in the form
\begin{equation}
  G(\ell)   = -{1\over M^2} \Big(\rlap/{\mkern-1mu}\ell + M\Big)
\bigg (1 + {\ell^2\over M^2} + \cdots \bigg) \;.\label{eq:fexpn}
\end{equation}
In dimensional regularization
$\int {\rm d}^d \ell \, \ell^n = 0$ so the soft 
part vanishes. Alternatively, a direct calculation of the loop integrals 
with the standard Feynman rules can be used to show that the contributions 
of the diagrams can be expanded in a power series in terms of small pion
momenta (a $Q/M$ expansion) and can thus be absorbed in the lagrangian. 
In other words there is no soft part.

\begin{figure}
 \setlength{\epsfxsize}{3in}
  \centerline{\epsffile{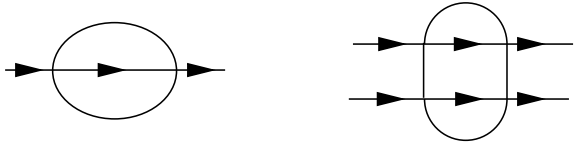}}
\vspace*{.2in}
\caption{Sample diagrams with baryon loops involving four
baryon vertices.}
 \label{fig:three}
\vspace*{.4in}
\end{figure}

\begin{figure}
 \setlength{\epsfxsize}{3in}
  \centerline{\epsffile{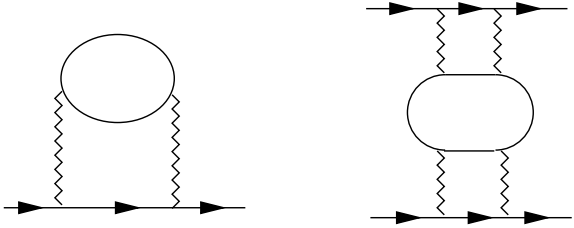}}
\vspace*{.2in}
\caption{Diagrams which yield Fig.~3 when the heavy boson propagators, 
represented by wiggly lines, are shrunk to a point.}
 \label{fig:four}
\vspace*{.4in}
\end{figure}

Next we discuss fermion loops generated from four (or more) fermion
vertices, such as those in Fig.~\ref{fig:three}.  
These can be considered to arise from Fig.~\ref{fig:four}
by shrinking the heavy boson lines to points.
Since the momentum transfers through the boson
lines must be small if it is feasible to integrate out
the heavy bosons, the arguments given in connection with
Fig.~\ref{fig:two} can also be applied here. Therefore the soft
parts vanish. Alternatively, this may be shown directly by 
taking the loop momenta to be of ${\cal O}(Q)$, using 
Eqs.~(\ref{eq:expN}) and (\ref{eq:fexpn}), together with the relation
\begin{equation}
     \int\!{\rm d}^d{\ell}
           {  (\ell^2)^m
          \over (2k\cdot \ell+ \alpha + i\epsilon)^n } = 0
                          \;,\label{eq:Ik}
\end{equation}
where $\alpha$ is independant of $\ell$.
Equation (\ref{eq:Ik}) is valid in dimensional regularization for 
any integers $m$ and $n$ (see Ref.~\cite{COLLINS84} for example).

\begin{figure}
 \setlength{\epsfxsize}{3in}
  \centerline{\epsffile{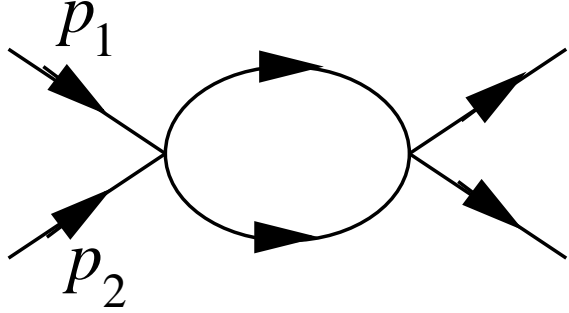}}
\vspace*{.2in}
\caption{ A two-baryon reducible diagram that needs special treatment.}
 \label{fig:five}
\vspace*{.4in}
\end{figure}

Finally, we need to consider the special case where an $N$-baryon 
scattering process contains an $N$-baryon intermediate state.
A simple example is given
in Fig.~\ref{fig:five}. We note that if we consider the
four-fermion interaction in terms of the exchange of a heavy 
boson, as before, this diagram will not contain a baryon loop. As
Weinberg noticed\cite{WEINBERG90}, 
there is an infrared divergence  in Fig.~\ref{fig:five}
for on-shell nucleons at zero kinetic energy. 
Indeed, the amplitude is proportional to
\begin{eqnarray}
 \int\! {\rm d}^4 \ell\,   P_1(\ell) G(p_1 + \ell) G(p_2 -\ell) 
          & =&   \int\! {\rm d}^4 \ell\,
      {  P_2(\ell)  \over   2p_1\cdot \ell+ p_1^2 - M^2 + i\epsilon} 
                         \nonumber  \\
        & & \quad \quad \quad 
            \times {1\over 2p_2\cdot \ell+ p_2^2 - M^2 - i\epsilon }
          + \cdots       \ ,
\end{eqnarray}
where $P_1(\ell)$ and $P_2(\ell)$ are polynomials in the loop 
momentum $\ell$. We have taken $\ell$ to be of order $Q$ and 
expanded the integrand in the manner of
Eq.~(\ref{eq:expN}). The contour of integration
is pinched between the two poles at $\ell_0 = \pm i\epsilon $ for
$p_1 = p_2 = (M, \bbox{0})$, and so cannot be distorted to avoid
these singularities. Of course, this just signals that
our expansion fails. The way out of this difficulty has also been 
given by Weinberg: we should consider only $N$-baryon
irreducible diagrams for $N$-baryon scattering processes.
The reducible diagrams are summed up with the 
$N$-body Bethe-Salpeter equation, the kernel of which is obtained
from the irreducible diagrams.

Even in one-baryon processes, such as $\pi N$ scattering, singular 
behavior can arise when an intermediate
$\Delta$ goes on shell. A similar remedy is followed:
first calculate the irreducible  self-energy diagrams to a 
certain order, then 
sum up the string of reducible diagrams containing arbitary 
numbers of self-energy insertions (see Sec. IV).

We can now discuss the power counting for irreducible diagrams
that do not contain fermion loops. According to 
Eq.~(\ref{eq:expN}) the leading order of a baryon propagator is
$Q^{-1}$ and according to Rule 1
the loop momentum is of order $Q$. Thus all the 
power-counting arguments of Weinberg \cite{WEINBERG90}
carry over. It follows
that the leading order of a Feynman diagram with $L$ loops, 
$E_{\rm N}$ external baryon lines
is $Q^\nu$ with
\begin{equation}
      \nu = 2+2 L -\case{1}{2}E_{\rm N} 
          + \sum_{i} V_i \Big (d_i + \case{1}{2} n_i -2\Big) 
                 \ , \label{eq:chcnt}
\end{equation}
where $V_i$ is the number of
vertices of type $i$ characterized by $n_i$ baryon
fields and $d_i$ pion derivatives or $m_\pi$ factors. 
(We used the quantity $d_i + \case{1}{2} n_i$ to characterize terms 
in the lagrangian in Sec. II.)
In general each diagram may contribute at orders beyond the leading
$Q^\nu$ order.

\section{Pion--Nucleon Scattering}

\begin{figure}
 \setlength{\epsfxsize}{3in}
  \centerline{\epsffile{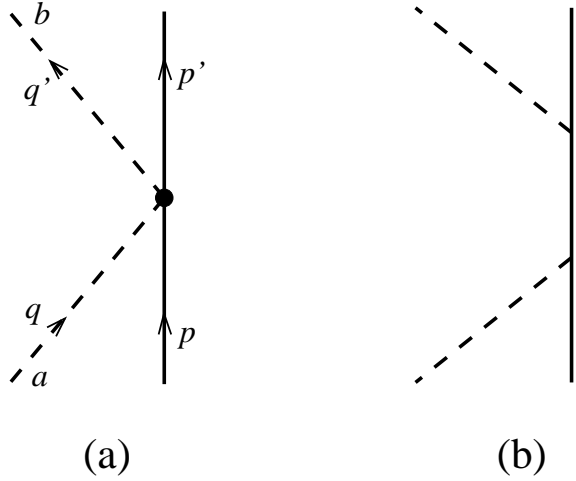}}
\vspace*{.2in}
\caption{Tree-level diagrams for $\pi N$ scattering. 
 (a) contact interactions; (b) nucleon exchange with the cross diagram 
suppressed.}
 \label{fig:treeandN}
\vspace*{.4in}
\end{figure}

We apply our formalism to $\pi N$ scattering 
and calculate the $T$ matrix
to \ord{3}. The $Q^3$ amplitude is obtained from tree diagrams 
constructed from
our lagrangian ${\cal L}_2 + {\cal L}_3 + {\cal L}_4$
and one-loop diagrams constructed from ${\cal L}_2$.
Following the 
standard notation of H\"{o}hler \cite{HOHLER83} and Ericson and
Weise \cite{EW88} we  write  the $T$ matrix as
\begin{equation}
  T_{ba} \equiv \langle \pi_b | T | \pi_a\rangle
         =T^+ \delta_{ab} + \case{1}{2}[\tau_b, \tau_a] T^- \ ,
\end{equation}
where the isospin symmetric and antisymmetric amplitudes are 
\begin{equation}
    T^{\pm} = A^{\pm} + \case{1}{2}(\rlap / q +
                           \rlap/ q') B^\pm \ .
\end{equation}
Here, as shown in Fig.~\ref{fig:treeandN}(a), $q$ and $q'$ are 
the c.m. momenta of the incoming and 
outgoing pions with isospin labels $a$ and $b$ respectively. The
c.m. momenta of the incoming and outgoing nucleons are labelled
$p$ and $p'$ respectively. The amplitudes
$A^\pm$ and $B^\pm$ are functions of the Mandelstam 
invariants $s=(p+q)^2$, $t=(q-q')^2$, and $u=(p-q')^2$.

\subsection{Tree-level Contact Terms \& Nucleon Exchange}

In Fig.~\ref{fig:treeandN} we show the tree level Feynman diagrams
arising from the contact terms and from one nucleon exchange, with
the crossed diagram for the latter suppressed.
The vertex in Fig.~\ref{fig:treeandN}(a) arises from any of the 
interactions in ${\cal L}_3$ and ${\cal L}_4$ (except for the 
$\lambda_1$ and $\lambda_6$ terms), as well as the Weinberg term
$-\overline{N}\gamma^{\mu}v_{\mu}N$. It is straightforward to obtain
the amplitudes arising from the contact terms. The results  are
\begin{eqnarray}
  A_{\rm C}^+ & = & {1\over M f_\pi^2}
    \left[ 2\beta_\pi\Big(2m_\pi^2 -t\Big) 
                 -4\kappa_2 m_\pi^2 
                 +\frac{\lambda_4}{8M^2} (s-u)^2 \right]  \ ,
                          \\
  B_{\rm C}^+ & = & {1\over 4M^2 f_\pi^2}(\kappa_1 - 2\lambda_4)(s-u)\ ,
                          \\
  A_{\rm C}^- & = & -{ \kappa_\pi \over  2Mf_\pi^2} (s-u)
                        \ ,         \\
  B_{\rm C}^- & = & {1\over 2 f_\pi^2}(1 + 4\kappa_\pi ) 
                   - { 1 \over  M^2 f_\pi^2}
                       \left[\case{1}{2}\lambda_2 t 
                             +4 \lambda_3 m_\pi^2 
                 -\frac{\lambda_5}{16M^2} (s-u)^2 
                        \right] \ .
\end{eqnarray}
The parameters here will absorb the divergences arising from 
the one-loop diagrams. They depend on the scale of 
dimensional regularization, $\mu$, in such a 
way that the complete $T$-matrix is $\mu$-independent.
The contributions from
nucleon exchange shown in Fig.~\ref{fig:treeandN}(b) are well-known, 
see for example Ref.~\cite{HOHLER83}. Including the crossed diagrams, 
we have
\begin{eqnarray}
  A_{\rm N}^+ & = & { g_{\rm A}^2\over f_\pi^2} M  \ ,
                          \\
  B_{\rm N}^+ & = & {  g_{\rm A}^2\over f_\pi^2} M^2
           \bigg ( {1\over u-M^2} - {1\over s-M^2} \bigg ) \ ,
                          \\
  A_{\rm N}^- & = & 0  \ ,
                          \\
  B_{\rm N}^- & = & -  {g_{\rm A}^2\over 2 f_\pi^2}
                 -{  g_{\rm A}^2\over f_\pi^2} M^2
                  \bigg ( {1\over s-M^2} +  {1\over u-M^2} \bigg ) \ .
\end{eqnarray}
Here we have used the exact nucleon propagator, although it could be 
expanded in chiral orders as in Ref.~\cite{MO97}. The difference would
appear beyond \ord{3} which is the level of precision of the present
calculation. Note that identification of what is to be included to      
\ord{3} is model dependent since the hard momentum scale could be       
$M$, $M_\Delta$, the average mass $\bar{M}$, or $4\pi f_\pi$. Results   
obtained with different choices will differ at \ord{4} and beyond, but  
this should not affect the quality of the fit to the data so that we    
shall simply write our expressions in convenient form.                  

\subsection{$\Delta$ Exchange}

\begin{figure}
 \setlength{\epsfxsize}{3in}
  \centerline{\epsffile{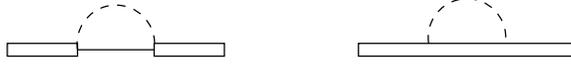}}
\vspace*{.2in}
\caption{One-loop $\Delta$ self-energy diagrams. The 
open box represents the free $\Delta$ propagator.}
 \label{fig:seven}
\vspace*{.4in}
\end{figure}

When the $\Delta$ appears as an intermediate state for $\pi N$ scattering, 
the tree-level $T$-matrix diverges at $s = M_\Delta^2$ so the power 
counting fails. As argued earlier, we expect 
the power counting to work only for irreducible diagrams.
Thus we evaluate the one-particle irreducible
self-energy diagrams which, to one-loop order, are those of  
Fig.~\ref{fig:seven}. Diagrams containing one or more self-energy 
insertions are then summed to replace the free propagator by the
dressed propagator which is finite.

We start by writing down the free $\Delta$ propagator
\begin{equation}
  G^0_{\mu\nu}(k) = {1\over  \rlap/{\mkern-2mu} k  - M_\Delta+i\epsilon}
        \left(-g_{\mu\nu} +\toneth\gamma_\mu\gamma_\nu
              + {1\over 3 M_\Delta}(\gamma_\mu k_\nu-
             \gamma_\nu k_\mu) +
             {2k_\mu k_\nu \over 3 M_\Delta^2} \right)
                  \ .\label{delfree1}
\end{equation}
This may be recast in 
terms of the spin projection operators \cite{NIEU81,BEN89} as
\begin{equation}
 G^0_{\mu\nu}(k) = -{1\over  \rlap/{\mkern-2mu} k  - M_\Delta+i\epsilon}
             (P^{3/2})_{\mu\nu}-\frac{1}{\sqrt{3}M_{\Delta}}
             (P^{1/2}_{12}+P^{1/2}_{21})_{\mu\nu}
             +\frac{2}{3M_{\Delta}^2}(\rlap/{\mkern-2mu} k +M_{\Delta})
             (P^{1/2}_{22})_{\mu\nu}\;,\label{delfree2}
\end{equation}
where
\begin{eqnarray}
  (P^{3/2})_{\mu\nu} &=& g_{\mu\nu}- \toneth\gamma_\mu\gamma_\nu
         +\frac{1}{3k^2}(\gamma_\mu k_\nu-k_\mu\gamma_\nu)\rlap/{\mkern-2mu} k
         -\frac{2}{3k^2}k_\mu k_\nu \ , \nonumber \\
  (P^{1/2}_{11})_{\mu\nu}&=&\toneth\gamma_\mu\gamma_\nu
         -\frac{1}{3k^2}(\gamma_\mu k_\nu-k_\mu\gamma_\nu)\rlap/{\mkern-2mu} k
         -\frac{1}{3k^2}k_\mu k_\nu\;,\nonumber\\
  (P^{1/2}_{12})_{\mu\nu}&=&\frac{1}{\sqrt{3}k^2}(-k_\mu k_\nu 
          +\gamma_\mu k_\nu\rlap/{\mkern-2mu} k)\;,\nonumber\\
  (P^{1/2}_{21})_{\mu\nu}&=&\frac{1}{\sqrt{3}k^2}(k_\mu k_\nu 
           -k_\mu\gamma_\nu \rlap/{\mkern-2mu} k)\;,\nonumber\\
  (P^{1/2}_{22})_{\mu\nu}&=&\frac{1}{k^2}k_\mu k_\nu\;.
\end{eqnarray}
The spin projection operators obey the orthogonality relations
\begin{equation}
  \left(P^I_{ij}\right)_{\mu\nu}\left(P^J_{kl}\right)^{\nu\rho}=
  \delta_{IJ}\delta_{jk}\left(P^I_{il}\right)_{\!\mu}^{\:\rho}\;.
\end{equation}

The dressed $\Delta$ propagator contains any number
of irreducible self-energy insertions:
\begin{equation}
G_{\mu\nu} = G^0_{\mu\nu} 
                 +  G^0_{\mu\lambda}\Sigma^{\lambda\sigma}
                      G^0_{\sigma\nu} 
                 +  G^0_{\mu\lambda}\Sigma^{\lambda\sigma}
                    G^0_{\sigma\tau}\Sigma^{\tau\pi}
                    G^0_{\pi\nu} +\cdots 
               \ , \label{eq:Gmunu}
\end{equation}
where $\Sigma_{\mu\nu}(k)$ is the $\Delta$ self-energy. Since 
$\gamma^\mu G^0_{\mu\nu}(k)$ and $k^\mu G^0_{\mu\nu}(k)$ 
do not contain a pole at $k^2 = M_\Delta^2$, we conclude from 
Eq.~(\ref{eq:Gmunu}) that the tensor terms
in $\Sigma_{\mu\nu}(k)$ constructed from  $\gamma^\mu$ and $k^\mu$ 
generate non-pole terms in the dressed propagator $G_{\mu\nu}$. 
It is not hard to see that these terms start to contribute to 
the $T$-matrix  at order $Q^5$.
We can thus greatly simplify our calculations by noting that
in the $\Delta$ self-energy tensor
\begin{equation}
  \Sigma_{\mu\nu}(k) \equiv \Sigma_\Delta(k) g_{\mu\nu} + \cdots \ ,
\end{equation}
the aforementioned tensor terms, represented by the ellipsis, can be 
neglected. Renormalizing $\Sigma_\Delta$ in the $\overline{\rm MS}$ 
scheme we obtain
$\Sigma_\Delta^{\overline{\rm MS}}$. We make additional on-shell 
mass and wavefunction counterterm subtractions 
such that, when the imaginary part of the self-energy is neglected, the 
pole of the $\Delta$ propagator lies at the physical mass with unit 
residue. Thus, the final renormalized self-energy is
\begin{equation}
    \Sigma_\Delta^{\rm ren}(k) =\Sigma_\Delta^{\overline{\rm MS}}(k) 
          -\Re\Sigma_\Delta^{\overline{\rm MS}}(k)
               \Bigm|_{\rlap/{\mkern-2mu} k=M_\Delta}
           -{\partial \over \partial  \rlap/{\mkern-2mu} k }
          \Re\Sigma_\Delta^{\overline{\rm MS}}(k)
            \Bigm|_{\rlap/{\mkern-2mu} k=M_\Delta}
             (\rlap/{\mkern-2mu} k-M_\Delta)\;,
\end{equation}
where $\Re$ refers to the real part.
Breaking the $\Delta$ self-energy into real and imaginary parts
we obtain the dressed $\Delta$ propagator
\begin{eqnarray}
G_{\mu\nu}(k) &=& -{\rlap/{\mkern-2mu} k + M_\Delta
                   \over k^2-M_\Delta^2 -\Pi_\Delta(\eta_k) 
              + i M_{\Delta}\Gamma_\Delta(k^2)} \, \,
              (P^{3/2})_{\mu\nu}
             -\frac{1}{\sqrt{3}M_{\Delta}}
             (P^{1/2}_{12}+P^{1/2}_{21})_{\mu\nu}\nonumber\\
&&\qquad\qquad+\frac{2}{3M_{\Delta}^2}(\rlap/{\mkern-2mu} k +M_{\Delta})
             (P^{1/2}_{22})_{\mu\nu} \ .
                      \label{eq:fullGmunu}
\end{eqnarray}
We have ignored \ord{3} contributions to the non-pole terms which 
involve $P^{1/2}$ since, as we have remarked, they do not 
contribute to the $\pi N$ $T$-matrix until \ord{5}.
Also in the numerator of the pole term in
Eq.~(\ref{eq:fullGmunu}) we have neglected 
terms of \ord{3}, which contribute to the $\pi N$ $T$-matrix at \ord{4}.
It is convenient to write the real part of the $\Delta$ ``polarization'' 
as a function of $\eta_k = \thalf (k^2 - M^2)/\bar{M}$, 
with the mean baryon mass $\bar{M} =\thalf (M + M_\Delta)$.
To \ord{3} the two diagrams of Fig.~\ref{fig:seven} yield
\begin{eqnarray}
\Pi_\Delta(\eta_k) & = & -{4\over 3} {h_A^2 M_\Delta\over (4\pi f_\pi)^2}
                    \bigg \{ \Big(\eta_k^2-m_\pi^2\Big) J(\eta_k)
                          -  \Big(\delta^2-m_\pi^2\Big) J(\delta)
                       \nonumber                  \\
         & &     \ \ \ \ \    \ \ \ \ \   \ \ \ \ \ \ \ \    
                   -\bigg[3\delta  J(\delta) -2 \delta^2 
                    +m_\pi^2 + m_\pi^2\ln {m_\pi^2\over \mu^2} \bigg]
                       (\eta_k -\delta) \bigg \}                        
                      \nonumber    \\
      & &   -{25\over 27} {\tilde{h}_A^2 M_\Delta\over (4\pi f_\pi)^2}
                   \bigg\{ \Big [ (\eta_k-\delta)^2-m_\pi^2 \Big]
                      J(\eta_k -\delta)
                   \nonumber  \\  
       & &       \ \ \ \ \  \ \ \ \ \ \ \ \ \ \ \ \ \ \ \ \
               - \pi m_\pi^3
                  -m_\pi^2\bigg(1 + \ln {m_\pi^2\over \mu^2} \bigg)
                       (\eta_k -\delta) \bigg\}
                    \ ,   
\end{eqnarray} 
where the function $J$ is defined in the Appendix.
For the energies considered here the imaginary part arises from the first 
diagram of Fig.~\ref{fig:seven} which gives
\begin{eqnarray}
   \Gamma_\Delta(k^2) &=& {\pi \over 12M_\Delta k^4}
                 {h_A^2 \over (4\pi f_\pi)^2}
                  \Big( k^2 +M^2 +2MM_\Delta\Big)
               \Big[ (k^2-M^2)^2-(k^2+3M^2)m_\pi^2 \Big] \nonumber\\
               &&\qquad\qquad\times\sqrt{(k^2-M^2)^2-4k^2m_\pi^2}\ .
\end{eqnarray} 
We have evaluated $\Gamma_\Delta$ up to \ord{4}
because this significantly improves the accuracy of the $\Delta$
decay width, $\Gamma_\Delta(k^2=M_\Delta^2)$. In fact the error
is negligible when compared with an exact evaluation
using tree-level coupling, $g_{\pi N\Delta}=h_AM/f_\pi$ (for the 
coupling to \ord{2} see Eq.~(\ref{eq:gpiNDel}) below). 
Specifically this is \cite{HOHLER83}
\begin{equation}
\Gamma_\Delta^{\rm exact}=\frac{g_{\pi N\Delta}^2}{12\pi}
\frac{|\vec{q}\:|^3}{M^2M_\Delta}\left(M+\sqrt{|\vec{q}\:|^2+M^2}\right)\;,
             \label{eq:exctwdth}
\end{equation}
where $|\vec{q}\:|^2=(4\bar{M}^2-m_\pi^2)(\delta^2-m_\pi^2)/(4M_\Delta^2)$.

\begin{figure}
 \setlength{\epsfxsize}{1in}
  \centerline{\epsffile{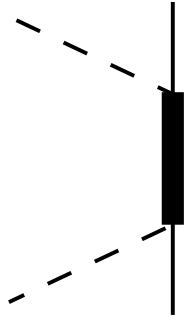}}
\vspace*{.2in}
\caption{The $\Delta$ exchange diagram with the dressed propagator
represented by a solid box.}
 \label{fig:delexch}
\vspace*{.4in}
\end{figure}

Using the dressed $\Delta$ propagator of Eq.~(\ref{eq:fullGmunu}), 
the $\Delta$ exchange contribution to the real part of the
$T$-matrix, pictured in Fig.~\ref{fig:delexch}, is
\begin{eqnarray}
  T_\Delta^{ba} &=& (\delta^{ab} -\tquar [\tau^b, \tau^a])\cdot
              { h^2_A \over 9M_\Delta^2f^2_\pi s} \,
         \biggl\{6M_\Delta^2\Delta_R(s)s[\alpha_1(s,t)+\alpha_2(s,t)
           \rlap/{\mkern0mu}q] \nonumber  \\
        & &-(s-M^2+m_\pi^2)\Bigl[M(s-M^2)+(M+2M_\Delta)m_\pi^2
     +(s-M^2+2MM_\Delta+m_\pi^2)\rlap/{\mkern0mu}q\Bigr]\!\biggr\}\,,
\end{eqnarray}
where we define $\eta\equiv \thalf (s-M^2)/\bar{M}$
and the real part of the $\Delta$ propagator is
\begin{equation}
 \Delta_R(s) = {s- M_\Delta^2 - \Pi_\Delta(\eta) \over
               [s- M_\Delta^2 - \Pi_\Delta(\eta)]^2 +
                    M_\Delta^2\Gamma_\Delta^2(s) }\;.
\end{equation}
We do not expand the propagator for the reason given at the end of 
Subsec. A.                                                         
The definitions of the functions $\alpha_1$ and $\alpha_2$ are
\begin{eqnarray}
    \alpha_1(s,t)& =& 2\bar{M}\Bigl[m_\pi^2-\thalf t-\toneth(s-M^2)
\Bigr]-\frac{(s-M^2+m_\pi^2)}{6s}
\Bigl[M(s-M^2)+m_\pi^2(M+2M_\Delta)\Bigr]\ ,    \nonumber \\
    \alpha_2(s,t)& = & \ttwoth m_\pi^2-\thalf t
+\tfourth M\bar{M}-\frac{(s-M^2+m_\pi^2)}{6s}
\Bigl(s-M^2+2MM_\Delta+m_\pi^2\Bigr) \ . \label{alph}
\end{eqnarray}
To this should be added the result for the cross diagram which 
is obtained from the above by the replacement $s\rightarrow u$
and the interchanges $a\leftrightarrow b$ and $q\leftrightarrow-q'$.

\begin{figure}
 \setlength{\epsfxsize}{5in}
  \centerline{\epsffile{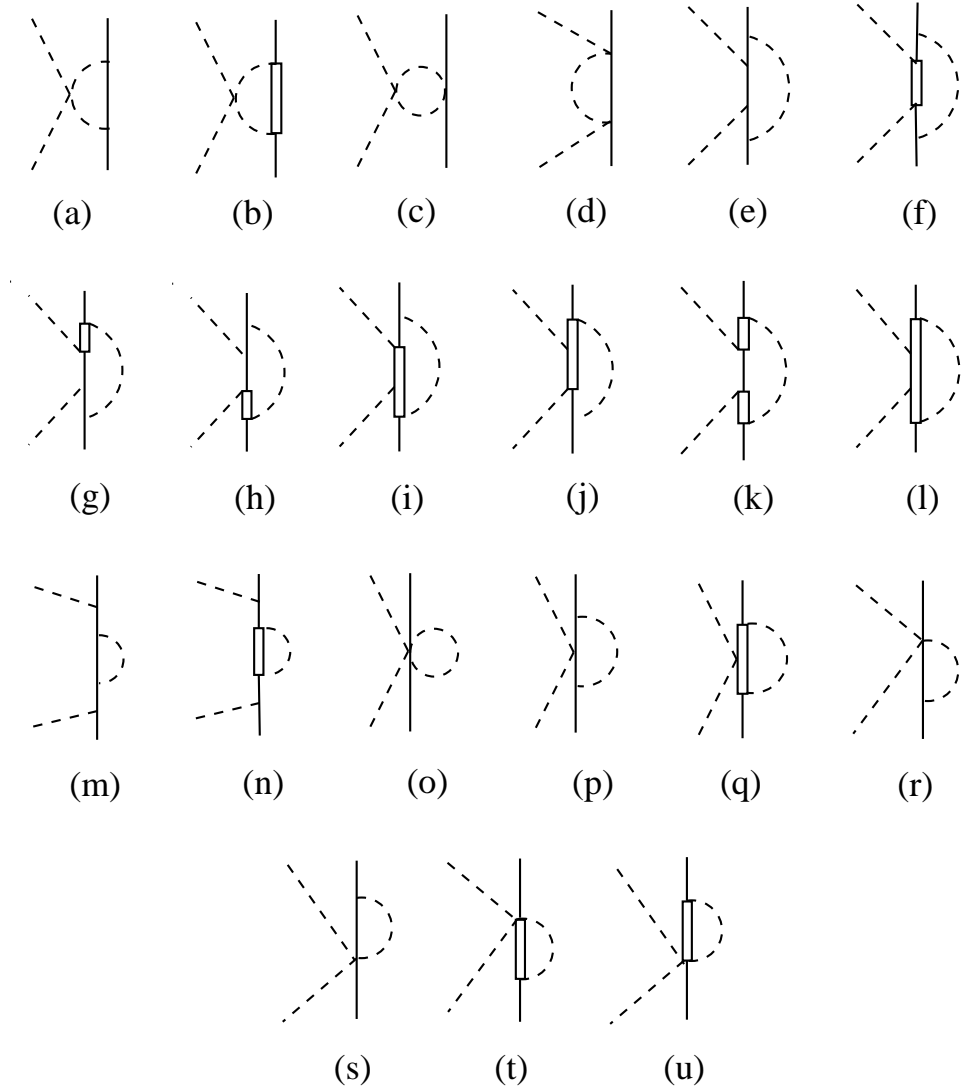}}
\vspace*{.2in}
\caption{A set of one-loop diagrams which contribute at \ord{3}. 
Crossed diagrams for (d) to (n) are not shown. }
 \label{fig:nonvtx}
\vspace*{.4in}
\end{figure}

\subsection{One-Loop Diagrams}

A set of one-loop diagrams that contribute to the $\pi N$ $T$-matrix at
\ord{3} is shown in Fig.~\ref{fig:nonvtx}. Here we can use
the free $\Delta$ propagator since no singularities are generated in    
the $T$-matrix. A covariant expansion of the $\Delta$ propagators, as   
well as the nucleon propagators, is                                     
made in the manner discussed in Sec. IIIA so that the denominators are  
of order $Q$. These $\Delta$ propagators are denoted by open boxes in   
Fig.~\ref{fig:nonvtx}.                                                  

\begin{figure}
 \setlength{\epsfxsize}{4.5in}
  \centerline{\epsffile{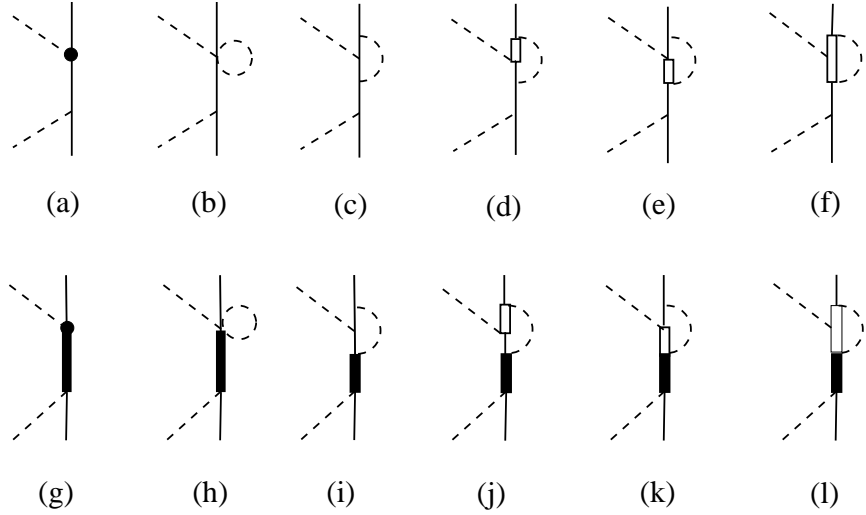}}
\vspace*{.4in}
 \caption{Diagrams with one-loop vertices which contribute at 
\ord{3}. Crossed diagrams are not shown. The solid circle in (a) and (g)
refers to $\lambda_1$ and $\lambda_6$ vertices, respectively.
Each diagram implicitly includes its counterpart
where the lower vertex is dressed.}
\vspace*{.4in}
 \label{fig:vtx}
\end{figure}

As an illustration, we calculate the $T$-matrix for 
Fig.~\ref{fig:nonvtx}(e). Using the standard Feynman rules, we find 
\begin{eqnarray}
   \overline{u}(p') T^{ba}_{9e} u(p) &=&
           -{g_A^4\over 16 f_\pi^4} \, i\mu^{4-d}\!
           \int\!{{\rm d}^d \ell\over ( 2\pi)^d}\,
              {1\over \ell^2-m_\pi^2 +i\epsilon}
                      \nonumber \\
           & & \ \ \ \ \ \ \ \  \times \overline{u}(p') 
              \rlap/{\mkern-1mu}\ell \gamma_5 \tau^c
              G(p'+\ell) \rlap/q' \gamma_5 \tau^b
              G(p+q+\ell)\rlap/q \gamma_5 \tau^a
              G(p+\ell)\rlap/{\mkern-1mu}\ell \gamma_5 \tau^c u(p)\ .
\end{eqnarray}
Noting that $p'_\mu=p_\mu+q_\mu-q'_\mu$, and that $q,\ q'$, and
$\ell$ are of order $Q$, we can expand the nucleon propagators as in 
Eq.~(\ref{eq:expN}) and select the leading terms. Then the 
contribution to \ord{3} is
\begin{eqnarray}
   \overline{u}(p') T^{ba}_{9e} u(p) &=&
           -{3g_A^4\over 16 f_\pi^4} \, i\mu^{4-d}\!
           \int\!{{\rm d}^d \ell\over ( 2\pi)^d}  \nonumber \\[5pt]
           & & \ \ \ \ \ \ \ \  \times
            {   \overline{u}(p') 
              \rlap/{\mkern-1mu}\ell (\rlap/{\mkern-1mu}p'-M)
              \rlap/q' 
              (\rlap/{\mkern-1mu}p + M)   
              \rlap/q    
              (\rlap/{\mkern-1mu}p - M)
              \rlap/{\mkern-1mu}\ell
              \Big(\delta^{ab} - \case{1}{6}[\tau^b,\tau^a]\Big) u(p)
                        \over (\ell^2-m_\pi^2 +i\epsilon)
        (2p\cdot \ell+i\epsilon)^2 (2p\cdot \ell +2p\cdot q+i\epsilon) }
\label{5e}
\end{eqnarray}
As in Sec. III, we can cast this in \hch\ form. Setting $p_\mu=Mv_\mu$
for the incoming nucleon, we easily obtain
\begin{eqnarray}
   \overline{u}(p') T^{ba}_{9e} u(p) &=&
           -{3g_A^4\over f_\pi^4} \, i\mu^{4-d}\!
           \int\!{{\rm d}^d \ell\over ( 2\pi)^d}  \nonumber \\[5pt]
           & & \ \ \ \ \ \ \ \  \times
           {\overline{u}(p')(\ell\cdot S)^2(q'\cdot S) (q\cdot S)
       \Big(\delta^{ab} - \case{1}{6}[\tau^b,\tau^a]\Big) u(p)  \over 
         (\ell^2-m_\pi^2 +i\epsilon)
        (v\cdot \ell+i\epsilon)^2 (v\cdot \ell +v\cdot q+i\epsilon) }\;.
\label{hb5e}
\end{eqnarray}
It is tedious, but straightforward to evaluate the numerators of 
Eqs.~(\ref{5e}) or (\ref{hb5e}) and carry out the integration using the 
results in the Appendix. We use the modified minimal subtraction scheme 
($\overline{\rm MS}$) to remove the singularities and the resulting 
real part of the
$T$-matrix is listed in Eq.~(\ref{eq:5e}). 
Note that divergent terms proportional to $1/(d-4)$ involves polynomials
in the variables and so they can be absorbed into the contact-term
contributions. For the same reason, we can absorb
finite polynomial terms obtained from 
the product of a $1/(d-4)$ singularity and 
$(d-4)$ factors. The latter may be generated from $d$-dimensional
$\gamma$-matrix algebra or from integration of tensor factors such as
$\ell_\mu \ell_\nu$ in the numerator of the integrand.
The above  simplification has been exercised for the results listed
in the Appendix.

The real parts of the $T$-matrix for the remaining diagrams of 
Fig.~\ref{fig:nonvtx} are evaluated in like fashion to \ord{3} and 
listed in the Appendix. Note that the diagrams in  
Fig.~\ref{fig:nonvtx}(m) and (n) involve the nucleon one-loop 
self-energy which is real for the energies of interest here. 
As with the $\Delta$, we renormalize in the $\overline{\rm MS}$ scheme 
to obtain $\Sigma_N^{\overline{\rm MS}}$ and make
additional on-shell mass and wavefunction counterterm 
subtractions. Thus, the renormalized self-energy is
\begin{equation}
 \Sigma_N^{\rm ren}(k) =\Sigma_N^{\overline{\rm MS}}(k) 
         -\Sigma_N^{\overline{\rm MS}}(k)\Bigm|_{\rlap/{\mkern-2mu} k=M}
           -{\partial \over \partial  \rlap/{\mkern-1mu} k }
               \Sigma_N^{\overline{\rm MS}}(k)\Bigm|_{\rlap/{\mkern-1mu} k=M}
                    (\rlap/{\mkern-2mu} k-M)\;. \label{eq:subnself}
\end{equation}
Since these diagrams do not give singular contributions to the $T$-matrix
we do not sum the self-energy insertions.

\begin{figure}
 \setlength{\epsfxsize}{5in}
  \centerline{\epsffile{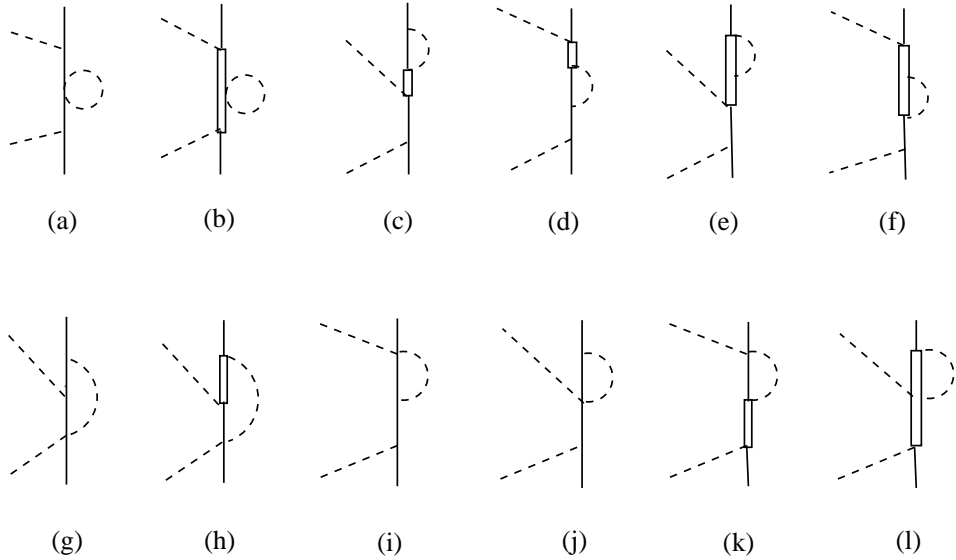}}
\vspace*{.2in}
\caption{One-loop diagrams which either vanish, (a) to (f),
or do not contribute at \ord{3}, (g) to (l). Crossed diagrams
and diagrams with the time ordering reversed are not explicitly 
shown.}
 \label{fig:zerodgms}
\vspace*{.4in}
\end{figure}

We also need to evaluate the vertex modification diagrams shown in 
Fig.~\ref{fig:vtx}. The $\Delta$ propagators denoted by open boxes     
have the same meaning as for Fig.~\ref{fig:nonvtx}. However there are  
now cases where the $\Delta$ can go on shell for which we use the      
dressed propagator of Eq.~(\ref{eq:fullGmunu}), as indicated by the    
solid box. For the diagrams in Fig.~\ref{fig:vtx}(i) and (j) 
the contributions proportional to 
$\Delta_R(s)$ vanish at the pole where 
$s=M_\Delta^2$ (and so $\eta=\delta$). As a result, at the pole there are 
important contributions to the $T$-matrix coming from the imaginary
part of the $\pi N \Delta$ vertex. These contributions actually
are of \ord{1} for energies very close to the $\Delta$ resonance
($|\eta-\delta| \ll m_\pi$). Thus, we keep the leading-order
pole corrections in Eqs.~(\ref{eq:10g}) and (\ref{eq:10h}).
We have also considered the next-to-leading order 
(\ord{2}) pole corrections for Fig.~\ref{fig:vtx}(i) and (j), as 
well as the leading \ord{3} pole correction to the $\Delta$ exchange diagram
of Fig.~\ref{fig:delexch}. These gave a very small effect and were not 
included in the final fit. The expressions for the vertex modification 
contributions to the $T$-matrix are also listed in the Appendix.
We reiterate that our 
expressions for diagrams that do not involve the $\Delta$
propagator agree with those recently given by \mo\ \cite{MO97}.
Finally we show in Fig.~\ref{fig:zerodgms} 
a set of one-loop diagrams which do not need to be considered. The top 
row, \ref{fig:zerodgms}(a)--(f), is identically zero due to the vertex 
structure or to violation of isospin conservation. The lower row,      
\ref{fig:zerodgms}(g)--(l), does not contribute to \ord{3}.            
This is a welcome simplification.                                      

From the real part of the $T$-matrix the real part of the elastic 
scattering amplitude, $f_\alpha$, is obtained by 
the standard partial wave expansion\cite{GASIO66}. 
Here the isospin-spin partial wave channels are labelled by
$\alpha\equiv(l,2I,2J)$ with $l$ the orbital angular momentum,
$I$ the total isospin, and $J=l\pm{1\over 2}$ the total angular momentum.
The phase shifts $\delta_\alpha$ are then given by 
\begin{equation}
     \Re f_\alpha = {1\over |\bbox{q}|}
             \sin \delta_\alpha \cos \delta_\alpha \;.   \label{eq:f}
\end{equation}

\subsection{Pion--Nucleon $\sigma$ Term}

We may obtain the nucleon $\sigma$ term from the Feynman-Hellman
theorem,
\begin{equation}
  \sigma(0) = m_\pi^2 {\partial M \over \partial m_\pi^2 } \ .
\end{equation}
Working out the nucleon self-energy up to one-loop $Q^3$ order, we find
\begin{eqnarray}
      M &=& M_0 -{4\kappa_2 \over M}m_\pi^2
            -{3\over 2} {\pi g_A^2 \over (4\pi f_\pi)^2} m_\pi^3 
                    \nonumber \\
           & & \ \ \ \ \ \ \ \  
           +{8\over 3} {h_A^2 \over (4\pi f_\pi)^2} 
            \bigg[
                      (\delta^2 - m_\pi^2) J(\delta)
                    + \thalf m_\pi^2 \delta 
                           \ln {m_\pi^2 \over \mu^2 }
               \bigg]   \ ,
\end{eqnarray}
where the function $J$ is defined in the Appendix and $M_0$ is the 
``bare" nucleon mass defined to be independent of the pion mass.
Note that, as in \hch, contact terms proportional to $\delta^3$ are
needed to absorb the divergences and $\mu$ dependence associated with
the $\Delta$ degrees of freedom. We have not specified these explicitly, 
hence our $M_0$ depends on $\delta$ and $\mu$.
The $\sigma$ term up to $O(Q^3)$ is then
\begin{equation}
 \sigma(0) = 
         -{4\kappa_2 \over M}m_\pi^2
            -{9\over 4} {\pi g_A^2 \over (4\pi f_\pi)^2} m_\pi^3
           +{8\over 3} {h_A^2 m_\pi^2 \over (4\pi f_\pi)^2}
    \Big[   \delta - \case{3}{2} J(\delta) \Big ]  \ .\label{eq:sigma}
\end{equation}
Note that the $\mu$ dependence in $J(\delta)$ may be absorbed in
$\kappa_2$ so that $\sigma$ is independent of $\mu$.

The isospin even on-shell amplitude $\overline{D}^+(\nu, t)$ 
at the Cheng--Dashen point can be related to the $\pi N$ $\sigma$ term,
$\sigma(2m_\pi^2)$, as follows \cite{BROWN71}:
\begin{equation}
f_\pi^2 \overline{D}^+(\nu=0, t=2m_\pi^2)
       = \sigma(2m_\pi^2) + {\cal O}(m_\pi^4) \ ,       
\end{equation}
where $\nu=(s-u)/(4M)$ and $\overline{D}^+$ is obtained from 
\begin{equation}
 D^+(\nu, t)  = A^+(\nu, t) + \nu B^+(\nu, t) 
\end{equation}
by the subtraction of the nucleon pole term which comes from the
tree diagram in Fig.~\ref{fig:treeandN}(b) and the one-loop
diagrams in Figs.~\ref{fig:vtx}(a)  to \ref{fig:vtx}(f).
At  the Cheng--Dashen point only Figs.~\ref{fig:delexch},
\ref{fig:nonvtx}(a), \ref{fig:nonvtx}(b)
and Figs.~\ref{fig:nonvtx}(r) to \ref{fig:nonvtx}(u) 
contribute to the $\sigma$ term. Using the results for the $T$-matrix
as given in the Appendix, we find
\begin{eqnarray}
  \sigma(2m_\pi^2) -  \sigma(0)      &=& 
            {3\over 4} {\pi g_A^2 \over (4\pi f_\pi)^2} m_\pi^3
            +  {h_A^2 m_\pi^4 \over 3 M_\Delta^2\delta}  \nonumber \\
           &+ &  
             {2\over 3} {h_A^2m_\pi^2 \over (4\pi f_\pi)^2}
                  \left[
                         (\pi - 4) \delta
                        - 4 \sqrt{\delta^2-m_\pi^2}
                           \ln  \left(
                            {  \delta\over m_\pi}
                         + \sqrt{ {\delta^2\over m_\pi^2} - 1} \,\right)
                 \right.   \nonumber \\
           &+ &    \left.
                \int_0^1\!{\rm d}x 
                     { 2\delta^2\over
                         \sqrt{\delta^2 - m_\pi^2(1-2x+2x^2)} }
                 \ln {\delta+  \sqrt{\delta^2 - m_\pi^2(1-2x+2x^2)}
                      \over \delta- \sqrt{\delta^2 - m_\pi^2(1-2x+2x^2)} }    
             \,  \right]   \ . \label{sigt0}
\end{eqnarray}
Note that since we count $\delta$ to be of order $Q$ the second term 
on the right is of the same $Q^3$ order as the other terms.
Equation (\ref{sigt0}) agrees with the result of Bernard {\it et al.}
\cite{BERNARD95,BERNARD96b} who
obtain this expression from HBChPT and dispersion relations. (Their integral 
is in a form which is slightly different from ours, but both integrals
give the same numerical result.)

\subsection{$\pi NN$ and $\pi N \Delta$ Couplings}

The $\pi NN$ vertex up to one-loop order consists of the tree vertex generated
from the axial $a_\mu$ term in the lagrangian and 
the one-loop diagrams shown in the upper part of 
Figs.~\ref{fig:vtx}(a) to (f).
Following our procedure, we can straightforwardly calculate the 
one-loop vertex function $\Gamma^a(k,k',q)$, where $k$ ($k'$ ) is the 
incoming (outgoing) momentum of the nucleon and $q=k'-k$ is the momentum
transfer. The $\pi NN $ coupling for on-shell nucleons is obtained by
sandwiching this vertex function between nucleon spinors:
\begin{equation}
    \overline{u}(k') \Gamma^a(k,k',q) u(k)
          = g_{\pi NN}(q^2)  \overline{u}(k')\gamma_5 \tau^a u(k) \ .
\end{equation}
We find
\begin{eqnarray}
    g_{\pi NN}(q^2) = g_{\pi NN}(0) + {\cal O}(q^2Q^2) \ , \label{eq:gq2}
\end{eqnarray}
which implies that the difference between the on-shell pion coupling
$g_{\pi NN}(m_\pi^2)$ and $ g_{\pi NN}(0)$ is of \ord{4}. This justifies 
the usual assumption of a small variation in 
$g_{\pi NN}(q^2)$ between $q^2=0$ and $q^2=m_\pi^2$
when the Goldberger-Treiman relation is derived \cite{CHENG84}. 
To \ord{2} we obtain
\begin{eqnarray}
    g_{\pi NN}(0) &=& {M g_A \over f_\pi} \bigg\{
                1 - {2\lambda_1 \over g_A}  {m_\pi^2 \over M^2}
                  -{1\over 3} {m_\pi^2 \over (4\pi f_\pi)^2}
                     \ln {m_\pi^2 \over \mu^2}
               + {g_A^2 \over 6} {m_\pi^2 \over (4\pi f_\pi)^2}
                    \bigg(1+\thrhalf\ln {m_\pi^2 \over \mu^2} \bigg)
                               \nonumber \\
           & & \ \ \ \ \ \ \ \ \ \  \
               +{64\over 27\delta}{h_A^2\over (4\pi f_\pi)^2}
                 \bigg[ \pi m_\pi^3
                       +\Big(\delta^2 - m_\pi^2\Big)J(\delta)
                       +\thalf m_\pi^2 \delta \ln {m_\pi^2 \over \mu^2}
                  \bigg]  \nonumber \\
           & & \ \ \ \ \ \ \ \ \ \  \
               +{200\over 243 g_A}{h_A^2 \tilde{h}_A \over (4\pi f_\pi)^2}
                    \bigg[ 2\delta^2 - m_\pi^2
                       - 3 \delta J(\delta)
                       -\thrhalf m_\pi^2 \ln {m_\pi^2 \over \mu^2}\bigg ]
              +{\cal O}(Q^3)   \bigg\}    \ .  \label{eq:gpiNN}
\end{eqnarray}
Notice that $\lambda_1$ absorbs the divergences and $\mu$-dependence
arising from the one-loop vertices. With the parameters obtained 
from fits to the $\pi N$ phase shifts Eq.~(\ref{eq:gpiNN}) 
allows a test of the Goldberger-Treiman relation.

Similarly, we can calculate the $\pi N \Delta$ vertex
from the tree-level $h_A$ term
and the one-loop diagrams shown in the upper part of 
Figs.~\ref{fig:vtx}(g) to (l).
Here, in the vertex function $\Gamma^{\mu a}(k,k',q)$ the label
$k$ now refers to the incoming $\Delta$ momentum.
The $\pi N\Delta$ coupling is obtained by
sandwiching this vertex function between the nucleon spinor 
and the $\Delta$ spinor, $u_\mu(k)$:
\begin{equation}
    \overline{u}(k') \Gamma^{\mu a}(k,k',q) u_\mu (k)
          = g_{\pi N\Delta}(q^2)  \overline{u}(k')q^\mu T^a u_\mu (k) \ .
\end{equation}
We find
\begin{eqnarray}
    g_{\pi N\Delta}(q^2) = g_{\pi N\Delta}(0)
                 + {\cal O}(q^2Q^2) \ , \label{eq:gq2Del}
\end{eqnarray}
where
\begin{eqnarray}
    g_{\pi N\Delta}(0) &=& {M h_A \over f_\pi} \bigg\{
                1 - {2\lambda_6 \over h_A}  {m_\pi^2 \over M^2}
                  -{1\over 3} {m_\pi^2 \over (4\pi f_\pi)^2}
                     \ln {m_\pi^2 \over \mu^2} \nonumber \\
           & & \ \ \ \ \  \
               + {2\over 3\delta} {1\over (4\pi f_\pi)^2}
                  \Big(g_A^2 +\ton h_A^2+\ttfeo\tilde{h}_A^2
               \Big) \bigg[
                  \Big(\delta^2 - m_\pi^2\Big)J(\delta)
                      +\thalf  m_\pi^2  \delta \ln {m_\pi^2 \over \mu^2}
                \bigg]       \nonumber \\
           & & \ \ \ \ \ \          
        -{2\pi \over 3\delta}{1 \over (4\pi f_\pi)^2}
                    \Big(g_A^2 + \ttfeo\tilde{h}_A^2 \Big)
                              m_\pi^3 
               -{25\over 54}{g_A\tilde{h}_A m^2_\pi\over (4\pi f_\pi)^2}
                   \bigg(1+\thrhalf\ln {m_\pi^2 \over \mu^2} \bigg)
                            \nonumber \\
           & & \ \ \ \ \ \ 
               +{1\over 3 \delta}{2\pi i \over (4\pi f_\pi)^2}
               \Big(2g_A^2 + \ton h_A^2 \Big)
                    \Big( \delta^2 - m_\pi^2 \Big)^{{3\over 2}}
              +{\cal O}(Q^3)   \bigg\}    \ .  \label{eq:gpiNDel}
\end{eqnarray}
Here $\lambda_6$ plays a similar role to $\lambda_1$ for $g_{\pi NN}$.
The value of $ g_{\pi N\Delta}$ is complex because the 
intermediate pion and nucleon states for Fig.~\ref{fig:vtx}(i) 
and (j) can go on shell.

\section{Results}

\begin{figure}
 \setlength{\epsfxsize}{6.0in}
  \centerline{\epsffile{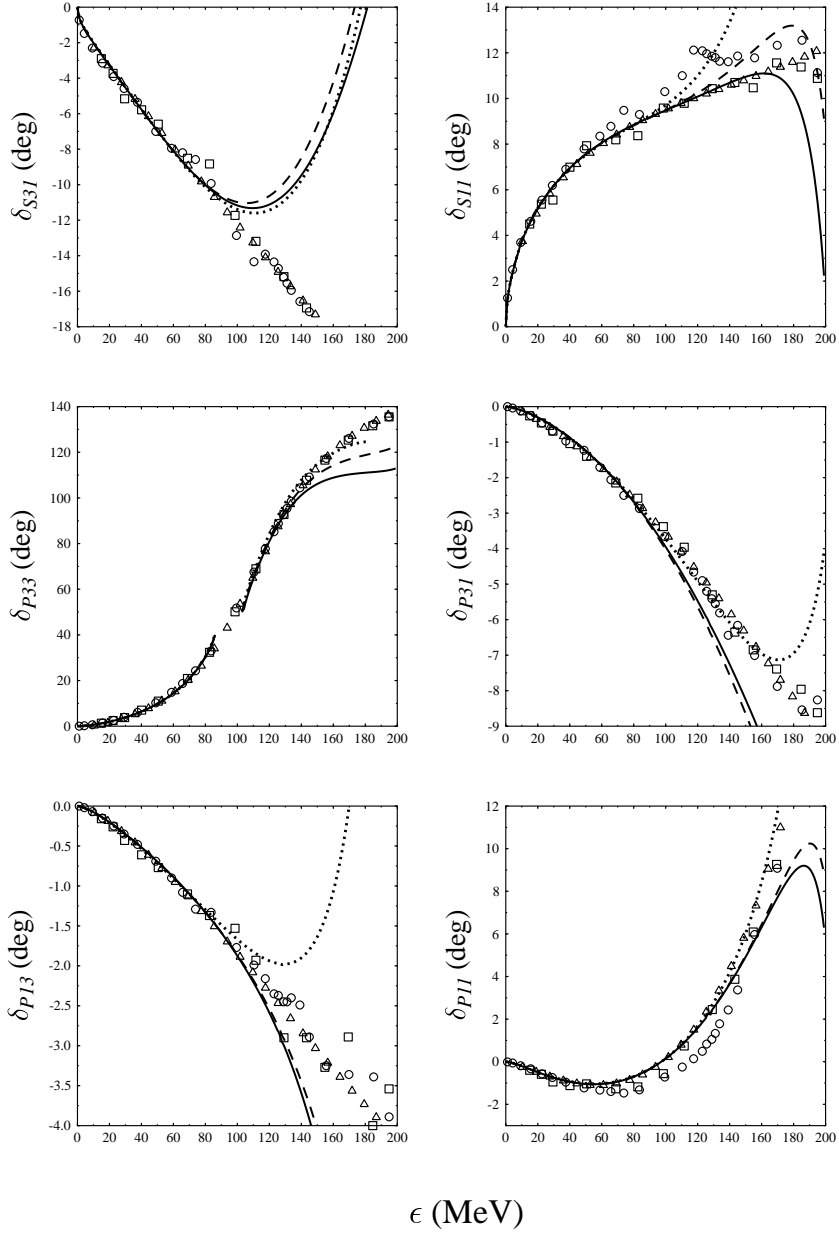}}
\vspace{-1.2in}
\caption{
The $S$- and $P$-wave phase shifts as a function of the pion c.m.
kinetic energy. The solid, dashed, and dotted curves are calculated with
parameters sets A1 ($\mu=1$ GeV), A2 ($\mu=1.232$ GeV), 
and A3 ($\mu=0.75$ GeV) respectively. The data are from 
Arndt\protect\cite{ARNDT}
(triangles), Bugg\protect\cite{BUGG} (squares), and Koch and
Pietarinen\protect\cite{KP80} (circles).}
 \label{fig:phshftmu}
\end{figure}

As fixed input parameters, we use the standard baryon and pion masses:
$M=939\,$MeV, $M_\Delta=1232\,$MeV, and $m_\pi=139\,$MeV.
We also take\cite{PDT} $f_\pi = 92.4\,$MeV from charged pion decay, 
$g_{\rm A} = 1.26$ from
neutron $\beta$ decay, and $h_{\rm A}=1.46$ from 
Eq.~(\ref{eq:exctwdth}) with $g_{\pi N\Delta} = h_A M / f_\pi$
and the central value
of the $\Delta$ width $\Gamma_\Delta = 120\pm5\,$MeV.
The physical results should be independent of 
the scale of dimensional regularization. To confirm that this is indeed the
case we carry out calculations with three values of the scale,
namely $\mu =M_\Delta$, $\mu = 1\,$GeV, and 
$\mu = 0.75\,$GeV. We then have eleven free parameters left:
$\beta_\pi$, $\kappa_\pi$,
$\kappa_1$, $\kappa_2$, $\tilde{h}_A$, and $\lambda_1$ to $\lambda_6$.
These are obtained by optimizing 
the fit of our calculated  $S$- and $P$-wave phase
shifts to the $\pi N$ scattering data of Arndt \cite{ARNDT}.
Because negligible error bars are given in the data at low energies,
we assign all the data points the same relative weight
in a least-squares fit. We fit the data for pion c.m. kinetic energies 
between $10$ and $100$ MeV. 

\def\mc#1{\multicolumn{1}{c}{$\quad #1$}}
\def\zz{\phantom{0}}

\begin{table}[t]
\caption{Parameter sets obtained 
from fitting the $\pi N$ phase shifts for various values of the 
renormalization scale $\mu$. The deduced $\pi NN$ and 
$\pi N\Delta$ couplings and the nucleon $\sigma$ term are also given. 
}
\vspace{.1in}
\begin{tabular}[t]{crrrrr}
                 & \mc{A1}   & \mc{A2}   & \mc{A3}    & \mc{B1}   & \mc{C1}   \\
$\mu({\rm GeV})$  & \mc{1.00} & \mc{1.232} & \mc{0.75} & \mc{1.00} 
& \mc{1.00} \\
\hline
$\beta_\pi$      & $-$3.5790 & $-$4.2084 & $-$1.9714  & $-$3.1385 & $-$3.1641 \\
$\kappa_\pi$     &    2.3219 &    3.5370 &    0.1993  &    1.9697 &    2.3169 \\
$\kappa_1 $      &    5.6578 &    6.5464 &    1.5485  &   4.6914  &    5.2765 \\
$\kappa_2 $      & $-$2.3683 & $-$2.5845 & $-$2.1417  & $-$1.9931$^\dagger$
& $-$1.6286$^\dagger$ \\
$\lambda_1$      & $-$2.6652 & $-$1.9338 & $-$4.2176  & $-$2.2593 & $-$9.7605 \\
$\lambda_2$      & $-$6.4822 & $-$4.5533 & $-$7.7712  & $-$8.2178 &$-$11.7498 \\
$\lambda_3$      &   11.0504 &   11.9228 &    4.6877  &   10.7125 &   15.2134 \\
$\lambda_4$      & $-$4.4304 & $-$2.5495 & $-$6.3775  & $-$6.1955 & $-$9.6040 \\
$\lambda_5$      &    4.9702 &    3.5669 &    4.1456  &    5.3523 &    8.0874 \\
$\lambda_6$      &    9.3759 &    9.2927 &    4.8363  &    7.3192 &    7.4099 \\
$\tilde{h}_A$    &    1.6243 &    1.4418 &    1.0986  &    1.6779 &    2.4626 \\
\hline\hline
$\sigma(0)({\rm MeV})$  & 106     &  109 &  108   &    75 &   45 \\
$g_{\pi NN}/({g_A M \over f_\pi})$
                 &    1.0445 &  1.0378   &    1.0409   &    1.0091  &    0.9585 \\
$|g_{\pi N\Delta}|/({h_A M \over f_\pi})$
                 &    0.9898 &  1.0134   &    1.0337   &    1.0532  &    1.1294 \\
\end{tabular}
\label{tab:one}
\noindent $^\dagger$ Fixed so as to obtain the listed value of $\sigma(0)$.
\end{table}

\begin{figure}
 \setlength{\epsfxsize}{3.0in}
 \epsffile{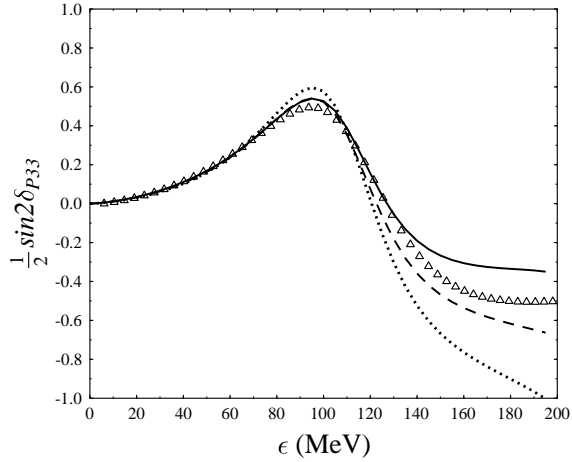}
\vspace{  .2in}
\caption{ 
The dependence of $\thalf\sin2\delta_{33}$ upon the pion c.m.
kinetic energy. The phase-shift data from
Arndt\protect\cite{ARNDT} are indicated by the triangles.
The solid, dashed, and dotted curves are calculated with
parameters sets A1, B1, and C1 respectively.}
 \label{fig:p33}
\end{figure}

The fit obtained for the $S$- and $P$-wave phase
shifts as  a function of the pion c.m. kinetic energy is indicated 
by the solid line in Fig.~\ref{fig:phshftmu}. Here the renormalization
scale is chosen to be 1 GeV and the corresponding
parameters are Set A1 in Table~1. The experimental data 
points of Arndt \cite{ARNDT} to which we fit are displayed as triangles
in Fig.~\ref{fig:phshftmu}; we also display there the older data of 
Bugg\cite{BUGG} (squares) and Koch and Pietarinen\cite{KP80} (circles).
The solid line shows a good fit up to 100~MeV pion c.m. kinetic energy,
slightly below the $\Delta$ resonance at 127 MeV.
It is remarkable that a good fit is achieved, even with eleven 
parameters, since the $T$-matrix contains a number 
of complicated non-polynomial functions of the invariant variables.
The agreement for most of the phase shifts extends beyond 100~MeV,
but $\delta_{S31}$ starts to deviate markedly.
Extending the range of c.m. energies used for the fitting does not
change the situation significantly. However,
\ord{4} contributions may become significant above 
approximately 100~MeV. This assertion is supported by test calculations
in which we have included a few selected contributions of this order.
However, without considering all \ord{4} contributions no 
definitive statement can be made. A better strategy
is to examine the renormalization scale dependence of the 
phase shifts. The dashed and dotted curves in Fig.~\ref{fig:phshftmu}
display results obtained with $\mu=M_\Delta$ and $\mu=0.75$ GeV 
respectively; the corresponding parameters are labelled Sets A2 and A3
in Table~1. The $\chi^2$ values for the three different values of the scale 
are similar and the phase shifts agree up to 100~MeV. Beyond this energy a
dependence on the scale starts to appear indicating that \ord{4} 
contributions are needed. The deviations 
are most noticeable for $\delta_{S31}$ and $\delta_{P13}$ 
which suggests that they are the most sensitive to higher-order 
contributions. We remark that similar fits can be obtained with values of 
the renormalization scale as small as $m_\pi$. In Fig.~\ref{fig:phshftmu}
there is a small gap in the $\delta_{P33}$ phase shift just below 100 MeV.
Here unitarity is slightly violated and a phase shift cannot be determined.
The point is made in Fig.~\ref{fig:p33} where 
$\thalf\sin2\delta_{P33}$
is plotted for Set A1 (solid curve). The maximum magnitude of $\thalf$ 
is exceeded in 
a 10 MeV energy region by $\leq 10$ \%. This is a small violation and the 
salient point is that sensible results are obtained without resorting to
a phenomenological $K$-matrix approach to enforce unitarity.

While we expect the parameters of Table~1 to be natural, {\it i.e.}
of order unity, 
some of them are close to 10. There could be several reasons for this.
First, we have no definite prescription for assigning
factors of 2 in the definition of the parameters. Second, the phase 
shifts are derived from the $T$-matrix where there are strong cancellations
among the various terms. This suggests that 
the parameters have significant uncertainties.
Third, we note that the parameters are scale dependent. They appear more 
natural for Set A3 with $\mu=0.75$ GeV than for Sets A1 and A2. 
An example of this is the parameter $\lambda_6$ 
which contributes to the effective $\pi N\Delta$ coupling,
see Eq. (\ref{eq:gpiNDel}). 
In fact the last row of the table shows that the effective 
$\pi N\Delta$ couplings obtained from sets A1, A2, and A3 
are quite similar and show a very small deviation from the tree-level 
value. Thus, we should not be unduly concerned by the size of some
of the parameters.

\begin{figure}
 \setlength{\epsfxsize}{6.0in}
  \centerline{\epsffile{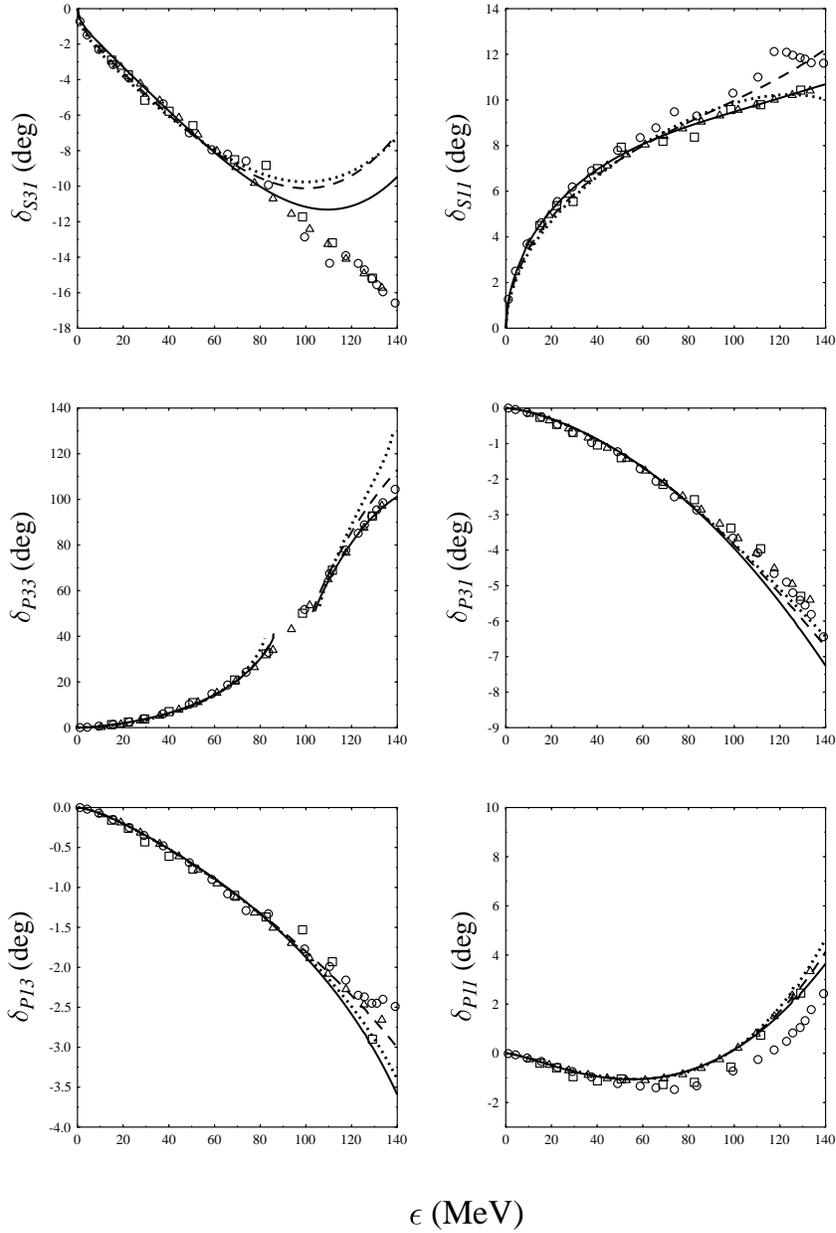}}
\vspace{-1.2in}
\caption{
The $S$- and $P$-wave phase shifts as a function of the pion c.m.
kinetic energy. The solid, dashed, and dotted curves are calculated with 
parameters sets A1, B1, and C1 respectively. The data are as in Fig. 12.}
 \label{fig:phshft}
\end{figure}

In Table~1 we also show the value of the effective $\pi NN$ 
coupling, which remains close to the tree-level value.
The $\pi NN$ coupling compares quite well with the determination of
Arndt {\it et al.} \cite{ARNDT94} of 
$g_{\pi NN}/({g_A M \over f_\pi})=1.03$.
We also tabulate the nucleon sigma term calculated from Eq.~(\ref{eq:sigma}).
Essentially the same value is obtained for Sets A1, A2, and A3 so the
sigma term is renormalization scale independent, as it should be.
The magnitude of 108 MeV, however, is
much larger than the generally accepted value of $45\pm 8$ due to
Gasser {\it et al.}\cite{GASSER91}. It is only slightly 
reduced to 92 MeV when we fit to the older data 
of Koch and Pietarinen\cite{KP80}. The value of 
$\sigma(2m_\pi^2)-\sigma(0)=15.7$ MeV depends only on $g_A$ and $h_A$
and is uncontroversial \cite{BERNARD95,BERNARD96,GASSER91}. 
Thus our predicted $\sigma(2m_\pi^2)$ is also much larger than the 
value of $64\pm8$ MeV obtained \cite{KOCH82} using dispersion relations. 
In order to see how sensitive our fit is to the sigma term,
we make two other fits in which 
$\sigma(0)$ is constrained to be 75 and 45 MeV respectively.
This is achieved by fixing $\kappa_2$ via Eq.~(\ref{eq:sigma}),
while allowing the remaining parameters to vary. We choose the scale of 
dimensional regularization to be $\mu=1\,$GeV. The parameters thus
obtained are respectively labelled B1 and C1 in Table~1.
The corresponding phase shifts are 
denoted by dashed and dotted lines in Fig.~\ref{fig:phshft} and
these can be compared with the solid curve obtained with Set A1.
The value of $\chi^2$ for Set B1 is 50\% larger than for Set A1
and only the $\delta_{S31}$ phase shift shows any visible difference 
between the two cases for c.m. energies below 100 MeV. The $\pi NN$ and 
$\pi N\Delta$ couplings are also similar. However, for set C1 $\chi^2$
is more than a factor of 2 greater than for Set A1 and larger
differences appear for the phase shifts, particularly the $S$-waves, 
and couplings.
Clearly $\sigma(0)$ is not well determined, although values larger
than 45 MeV are suggested. Note that \mo\ \cite{MO97}
obtained a value of 59 MeV by fitting to the threshold parameters,
the $\pi NN$ coupling, and the sigma term. The parameters for Sets 
A1, B1, and C1 show some significant differences indicating, 
as mentioned earlier, that they are 
not well determined by fitting the phase shift data alone.
We also note that Fig.~\ref{fig:p33} shows that Sets B1 and C1 violate 
unitarity at about 150 MeV c.m. energy, although here, as we have remarked,
we expect \ord{4} effects to be important. However, it may be significant 
that just below 100 MeV Set C1 gives a larger violation of unitarity 
than the other parameter sets.

\begin{table}[tbh]
\caption{The calculated $S$-wave scattering lengths and  $P$-wave
scattering  volumes, in units of $m_\pi^{-1}$ and $m_\pi^{-3}$
respectively, are compared with the data of Refs.~\protect\cite{ARNDT} 
and \protect\cite{KP80}.}
\vspace{.1in}
\begin{tabular}[t]{cddddd}
 length/volume & A1  & B1 & C1 & Ref.~\cite{ARNDT} &Ref.~\cite{KP80} \\
                        \hline
$a_1$    &    0.168 &    0.152 &    0.142 &    0.175 &    0.173$\pm$0.003  \\
$a_3$    & $-$0.080 & $-$0.095 & $-$0.105 & $-$0.087 & $-$0.101$\pm$0.004  \\
$a_{11}$ & $-$0.078 & $-$0.075 & $-$0.070 & $-$0.068 & $-$0.081$\pm$0.002  \\
$a_{13}$ & $-$0.026 & $-$0.026 & $-$0.025 & $-$0.022 & $-$0.030$\pm$0.002  \\
$a_{31}$ & $-$0.035 & $-$0.036 & $-$0.035 & $-$0.039 & $-$0.045$\pm$0.002  \\
$a_{33}$ &    0.201 &    0.188 &    0.171 &    0.209 &    0.214$\pm$0.002  \\
\end{tabular}
\label{tab:two}
\end{table}

It is also interesting to examine the threshold (vanishing pion 
kinetic energy) $S$-wave scattering lengths ($a_{2I}$)
and the $P$-wave scattering volumes ($a_{2I\,2J}$). 
The experimental values were not used in the fit and they are compared
with our predictions with Sets A1, B1, and C1 in Table~\ref{tab:two}.
The agreement for the $P$-waves is reasonable with not too much sensitivity 
to the parameter set, although $a_{33}$ does become  rather small for 
Set C1. For the $S$-waves it is instructive to examine the isoscalar and 
isovector scattering lengths,  $b_0=(a_1+2a_2)/3$ and 
$b_1=(a_3-a_1)/3$. For $b_1$,  
Ref.~\cite{ARNDT}, Ref.~\cite{KP80}, and a recent determination by 
Sigg {\it et al.} 
\cite{SIGG96} give $-0.091$, $-0.091\pm0.002$ and $-0.096\pm0.007$
respectively. These are in agreement and somewhat larger in magnitude than 
our predicted value of $-0.082$, virtually independent of parameter set.
The value of  $b_0$ from Ref.~\cite{ARNDT} is consistent with zero, while
Refs.~\cite{KP80,SIGG96} give $-0.010\pm0.003$ and $-0.008\pm0.007$. Our
parameter sets A1, B1, and C1 give 0.003, $-0.013$, and $-0.023$ 
respectively. From this result, and the others that we have discussed,
we conclude that parameter set C1 is less favored than the other sets.
Note that we do not show the scattering lengths and volumes for
Sets A2 and A3 because they differ negligibly from
Set A1, indicating $\mu$-independence. We make the obvious remark that 
the accuracy of the predicted scattering lengths and volumes can be    
improved by including them in the fit at the expense of a poorer fit   
to the phase shifts. In the case of Set A this reduces the sigma term  
by about 10\%.                                                         

\section{Conclusions}

We have introduced a new approach to chiral perturbation theory with 
baryons which involves manipulating the Feynman diagrams directly, 
rather than integrating out the heavy components of the baryon fields at
the lagrangian level as in \hch. Our approach preserves Weinberg's 
power counting so that a systematic expansion
in powers of a generic soft momentum scale, $Q$, remains valid.
We can maintain relativistic covariance and apply the standard Feynman rules
and $\gamma$-matrix algebra. At one-loop order we obtain equivalent
results to \hch; it would be interesting to compare the approaches
in higher order. 

We have calculated the $T$-matrix for \pin\ scattering up
to one-loop $Q^3$ order.
There were several components to this calculation.
First, the chiral lagrangian introduced a number of contact 
terms with unknown coefficients which contributed to the $T$-matrix.
Next, there were the well-known nucleon and $\Delta$ exchange diagrams.
The former was straightforward, but the latter was singular at tree
level when the $\Delta$ went on shell. The problem was solved 
by summing up \ord{3} self-energy insertions in the propagator to all 
orders. The resulting amplitude then (approximately) obeyed
unitarity for pion c.m. energies up to at least 140 MeV. Finally, we 
needed to evaluate the large number of one-loop
diagrams which could be constructed with the $\pi$, $N$ and $\Delta$ 
vertices. 

We have performed a least-squares fit
to the $S$- and $P$-wave phase shift data  to determine the parameters 
involved. We were able to obtain a good fit up to 100 MeV pion c.m. 
kinetic energy, slightly below the $\Delta$ resonance. 
Our predictions for the zero-energy scattering lengths
were also reasonable. These results were independent   
of the renormalization scale, $\mu$, as they should be.  
Since $\eta/\bar{M}$ is an expansion parameter 
and at the $\Delta$ resonance it becomes $\delta/\bar{M}=0.27$, one 
might expect \ord{4} effects to become important beyond the
$\Delta$ resonance energy.
This was supported by the scale dependence of our results in that
energy region. Though straightforward in principle, it would be very 
tedious to include the \ord{4} corrections.
The fit that we have obtained here contained an unavoidably large number of
parameters, but it must be borne in mind that the $T$-matrix is a much more
complicated function of the invariant variables when loops are present
than at tree level. Thus we regard the fit that we have achieved as
a vindication of the basic approach.
It seems, however, that one cannot determine the parameters with
great accuracy using the phase shift data alone.
In particular we were able to obtain very 
similar fits with parameters that corresponded to a nucleon $\sigma$ term
of 105 and 75 MeV. We judge that the fit was slightly degraded when
the currently accepted value of $\sigma(0)=45$ MeV was adopted, so 
that a somewhat larger value appears to be favored from the analysis here.
However this might indicate that \ord{4} effects need to be calculated 
in order to achieve an accurate one-loop result, as Ecker has recently 
argued \cite{ECKER97}.                                                 
It would be interesting to attempt to constrain the parameters further
and to compare with the parameters found at zero energy 
\cite{BERNARD96,MO97}. The latter would involve disentangling the $\Delta$
contributions as well as other effects that, for simplicity, we have 
absorbed in the contact terms.

 We thank S. Jeon and J. Kapusta for useful discussions and critical 
readings of the manuscript.
We acknowledge support from the Department of Energy under grant 
No. DE-FG02-87ER40328.

\appendix
\section*{}
We first list the integrals we need. Using dimensional regularization
we obtain
\begin{equation}
  i\mu^{4-d}\!\int\!{{\rm d}^d \ell\over ( 2\pi)^d}\,
    {1\over (\ell^2-m_\pi^2 +i\epsilon)[(\ell+Q)^2-m_\pi^2
      +i\epsilon]}=2L-\frac{1}{(4\pi)^2}I(Q^2)\;,\label{lint}
\end{equation}
where $L$, defined in Eq.~(\ref{eq:L}), is singular and the finite part
\begin{equation}
    I(x) = 1 - \ln {m_\pi^2 \over \mu^2 }
                - \sqrt{1 - {4m_\pi^2\over x}}
                      \,  \ln{ \sqrt{1 - {4m_\pi^2\over x}} + 1
                            \over
                            \sqrt{1 - {4m_\pi^2\over x}} - 1  }
\qquad{\rm for}\ x<0\;.
\end{equation}
The next integral to consider is
\begin{eqnarray}
 & & i\mu^{4-d}\!\int\!{{\rm d}^d \ell\over ( 2\pi)^d}\,
     {1\over (\ell^2-m_\pi^2 +i\epsilon)(2p\cdot \ell +z+i\epsilon) }
          =  \frac{z}{p^2}L-\frac{1}{(4\pi)^2\sqrt{p^2}}J\left(
         \frac{z}{2\sqrt{p^2}}\right)\nonumber\\
        &&\qquad\qquad\qquad\qquad-\frac{i}{16\pi p^2}\theta(z)
       \theta(z^2-4p^2m_\pi^2)\sqrt{z^2-4p^2m_\pi^2}\;,\label{jint}
\end{eqnarray}
where $\theta$ denotes the Heaviside step function and
\begin{eqnarray}
    J(x) &=&\left\{ 
                    \begin{array}{l}
                          x - x \ln {m_\pi^2 \over \mu^2 }
                                  - \sqrt{x^2 - m_\pi^2}
                      \,  \ln{ 1 +  {\sqrt{x^2 - m_\pi^2} \over x}
                             \over
                           1 -  {\sqrt{x^2 - m_\pi^2} \over x} }
                            \ \ \ \ \ \ \ \ \ \         (|x| > m_\pi) \\
                    x - x \ln {m_\pi^2 \over \mu^2 }
                     - 2  \sqrt{m_\pi^2 -x^2}\cos^{-1}
                       \Big( -{x\over m_\pi} \Big)      
                               \ \ \ \ \ \  \       (|x| < m_\pi)  
                    \end{array}
                \right.\;.
\end{eqnarray}
Note that $J(0)=-\pi m_\pi$.
The next integral we need is finite:
\begin{eqnarray}
&&i\mu^{4-d}\!\int\!{{\rm d}^d \ell\over ( 2\pi)^d}\,
{1\over (\ell^2-m_\pi^2 +i\epsilon)[(\ell+Q)^2-m_\pi^2 +i\epsilon]
        (2p\cdot \ell+i\epsilon)}=\frac{1}{16\pi\sqrt{p^2(4m_\pi^2-Q^2)}}
         S(Q^2)\;,\nonumber\\
&&\label{sint}
\end{eqnarray}
where
\begin{equation}
    S(x) =  \sqrt{1 - {4m_\pi^2\over x}}
                  \,  \sin^{-1} {1 \over
                         \sqrt{1 - {4m_\pi^2\over x} }}
\qquad\qquad{\rm for}\ x<0\;.
\end{equation}
Finally we require the finite integral
\begin{eqnarray}
&&i\mu^{4-d}\!\int\!{{\rm d}^d \ell\over ( 2\pi)^d}\,
{1\over (\ell^2-m_\pi^2 +i\epsilon)[(\ell+Q)^2-m_\pi^2 +i\epsilon]
        (2p\cdot \ell+M^2-M_\Delta^2+i\epsilon)}\nonumber\\
&&=\frac{1}{32\pi^2\bar{M}\delta}\int\limits_0^1\!dx\,K(x)\;,
\end{eqnarray}
where
\begin{eqnarray}
    K(x) &=& \left\{ 
              \begin{array}{l}
               {1 \over \sqrt{\delta^2 - m_\pi^2 +x(1-x) Q^2} }
               \, \ln {\delta+\sqrt{\delta^2 - m_\pi^2 +x(1-x) Q^2}
                \over 
               \delta-\sqrt{\delta^2 - m_\pi^2 +x(1-x) Q^2} } 
           \ \ \ \ \ \ \ \   ( \delta^2 - m_\pi^2 +x(1-x) Q^2 > 0 ) \\
            {2 \over \sqrt{m_\pi^2 -x(1-x) Q^2 - \delta^2}}
             \, \tan^{-1} {\sqrt{m_\pi^2 -x(1-x) Q^2 - \delta^2}
                  \over \delta }
           \ \ \ \ \ \    ( \delta^2 - m_\pi^2 +x(1-x) Q^2 < 0 )
                 \end{array}
           \right.\;.
\end{eqnarray}
In Ref. \cite{MO97} the expressions for the integrals of 
Eqs.~(\ref{lint}), (\ref{jint}), 
and (\ref{sint}) are written $-I_0(Q^2)$, 
$-J_0(z/\sqrt{4p^2})/\sqrt{4p^2}$, 
and $-K_0(0,Q^2)/\sqrt{4p^2}$; they agree with our results above. 
Our expressions are written in terms of $\bar{M}$, $\delta$, and 
$\eta$ and we recall the definitions for reference               
\begin{equation}                                                 
\bar{M}=\thalf(M+M_\Delta)\ ,\ \delta=M_\Delta-M\ ,              
\ \eta=\frac{s-M^2}{2\bar{M}}\ .                                 
\end{equation}                                                   
It is useful to define the functions
\begin{eqnarray}
 F_1(\eta,\delta) &=&  {1\over (\eta-\delta)^2}
                \bigg [ (\eta^2- m_\pi^2) J(\eta)
                       +(2\delta^2-3\eta\delta+ m_\pi^2)J(\delta)
                       \nonumber \\
           & & \ \ \   \ \ \ \ \ \ \ \ \ \   \ \ \ \ \ \ \
                    +(\eta-\delta)\bigg(2\delta^2- m_\pi^2
                           - m_\pi^2 \ln{m_\pi^2 \over \mu^2}
                             \bigg) \bigg ]
\\[5pt]
 F_2(\eta,\delta) &=&  {1\over \eta\delta(\eta-\delta)}
                \Big [ (\eta^2- m_\pi^2) \delta J(\eta)
                       -(\delta^2- m_\pi^2)\eta J(\delta)
                    +\pi m_\pi^3(\eta-\delta)\Big ]
\\[5pt]
 F_3(\eta,\delta) &=&  {1\over \eta-\delta}\,
                \bigg [ (\eta^2- m_\pi^2) J(\eta)
                       -(\delta^2- m_\pi^2) J(\delta)
                    +\thalf m_\pi^2 (\eta-\delta)
                     \ln {m_\pi^2 \over \mu^2 }\bigg ]\;.
\end{eqnarray}
These functions are finite, for example,
\begin{equation}
F_3(\eta,\delta)\stackrel{\eta\rightarrow\delta}{\rightarrow}
3\delta J(\delta)-2\delta^2+m_\pi^2+\thrhalf 
m_\pi^2\ln\frac{m_\pi^2}{\mu^2}\;.
\end{equation}

Then the contributions from Figs.~\ref{fig:nonvtx} and \ref{fig:vtx} 
to the real part of the $T$-matrix for $\pi N$ scattering are:
\begin{eqnarray}
  T^{+}_{9a} &=& {\pi g_A^2 \over 12f_\pi^2}{1\over (4\pi f_\pi)^2}
              \bigg[
                  (14m_\pi^2 - 12 t) m_\pi +
                     { 3(m_\pi^2 - 2 t) (2m_\pi^2 - t)
                        \over \sqrt{4m_\pi^2 - t}}S(t)  
               \bigg]
\;,\\[5pt]
  T^{-}_{9a} &=& {g_A^2 \over 12f_\pi^2} {1\over (4\pi f_\pi)^2}
              \bigg\{
                  - \eta(8m_\pi^2 - 5 t) I(t) +
                       10 \eta m_\pi^2 \ln {m_\pi^2 \over \mu^2 } 
                      \nonumber \\
          & & \ \ \ \ \ \ \ \ \ \ \ \
            -6\pi  \Big[ M \rlap / q -  \bar{M} \eta +
                          \case{1}{4}(2m_\pi^2 -t) \Big]  
                    \Big[2 m_\pi + \sqrt{4m_\pi^2 - t} \, S(t)\Big]  
               \bigg\}
\;,\\[5pt]
  T^{+}_{9b}  &=& {2 h_A^2 \over 9 f_\pi^2}{1\over (4\pi f_\pi)^2}
              \bigg[ \case{4}{3}
                     (7m_\pi^2 -6t -4 \delta^2)J(\delta)
                     +2\delta(m_\pi^2 -2t)I(t)  \nonumber \\
          & & \ \ \ \ \ \ \ \ \ \ \ \
                       -(m_\pi^2 - 2 t)(2 \delta^2 -2 m_\pi^2 +t)
                        \int_0^1\! {\rm d} x K(x)
                -\case{8}{3}  m_\pi^2\delta \ln {m_\pi^2 \over \mu^2 } 
               \bigg]
\;,\\[5pt]
  T^{-}_{9b}  &=& -{4 h_A^2 \over 9 f_\pi^2}{1\over (4\pi f_\pi)^2}
              \bigg\{\Big[ M \rlap / q -  \bar{M} \eta +
                          \case{1}{4}(2m_\pi^2 -t) +4\eta\delta\Big]J(\delta)
                             \nonumber \\
          & & \ \ \ \ \ \ \ 
                    +\Big[\Big( M \rlap / q -  \bar{M} \eta +
                          \case{1}{4}(2m_\pi^2 -t) \Big)\delta 
              +\Big(2\delta^2 -\case{4}{3}m_\pi^2+\case{5}{6}t\Big)\eta
                       \Big]I(t)       \nonumber \\
          & &\ \ \ \ \ \ \ 
              -\Big[\Big( M \rlap / q -  \bar{M} \eta +
                          \case{1}{4}(2m_\pi^2 -t) \Big)
                  \Big(\delta^2-m_\pi^2 +\case{1}{4} t\Big)
                                        \nonumber \\
          & &\ \ \ \ \ \ \ \ \ \ \ \ \ \ \ \ \
              + 2\eta\delta\Big(\delta^2-m_\pi^2 +\thalf t\Big)
                                  \Big]
                             \int_0^1\! {\rm d} x K(x)
              +\case{5} {3}  m_\pi^2\eta \ln {m_\pi^2 \over \mu^2 }
                                     \bigg\}
\;,\\[5pt]
  T_{9c}^{ba}  &=& - \thalf [\tau^b, \tau^a]\cdot
            {\eta \over 12 f_\pi^2}{1\over (4\pi f_\pi)^2}
              \bigg[
                     (4m_\pi^2 -t)I(t)
                        -2 m_\pi^2 \ln {m_\pi^2 \over \mu^2 } 
               \bigg]
\;,\\[5pt]
  T_{9d}^{ba}  &=&\Big (\delta^{ab} + \case{1}{4} [\tau^b, \tau^a]\Big)\cdot
                {\eta \over f_\pi^2}{1\over (4\pi f_\pi)^2}
              \bigg[  \eta J(\eta) 
                        +\case{3}{8} m_\pi^2 \ln {m_\pi^2 \over \mu^2 } 
               \bigg]
\;,\\[5pt]
   T^{ba}_{9e} &=&  -\Big(\delta^{ab} - \case{1}{6}[\tau^b,\tau^a]\Big)\cdot 
           {g_A^4\over 4 f_\pi^2} {1\over(4\pi f_\pi)^2 }
                     \Big ( M\rlap/q  - \bar{M}\eta
                  + 2m_\pi^2 - t -\case{3}{2}\eta^2 \Big) F_1(\eta,0)
                                         \label{eq:5e}
\;,\\[5pt]
   T^{ba}_{9f} &=&
             \Big(\delta^{ab} + \case{1}{12}[\tau^b,\tau^a]\Big)\cdot
           {2g_A^2\over 9 f_\pi^2}\, {h_A^2\over(4\pi f_\pi)^2 }
               F_1(\eta-\delta,0) \nonumber \\
           & & \  \ \ \ \ \
               \times \Big [  M\rlap/q  - \bar{M}\eta
                  - \case{5}{4}(2m_\pi^2 - t) +3\eta^2 \Big]
\;,\\[5pt]
   T^{ba}_{9g} &=&T^{ba}_{9h} = \case{1}{2}[\tau^b,\tau^a]\cdot
           {4g_A^2\over 27 f_\pi^2} {h_A^2\over(4\pi f_\pi)^2 }
            \Big[ M\rlap/q  - \bar{M}\eta
                  +\case{1}{4}(2m_\pi^2 - t) \Big]
              F_2(\eta,-\delta)
\;,\\[5pt]
   T^{ba}_{9i} &=&T^{ba}_{9j} = -\case{1}{2}[\tau^b,\tau^a]\cdot
           {20h_A^2\over 243 f_\pi^2} {g_A\tilde{h}_A\over(4\pi f_\pi)^2 }
           \Big[ M\rlap/q  - \bar{M}\eta
                  +\case{1}{4}(2m_\pi^2 - t) \Big]
                    F_2(\eta-\delta, -\delta)
\;,\\[5pt]
   T^{ba}_{9k} &=&\Big(\delta^{ab} +\case{5}{12}[\tau^b,\tau^a]\Big)\cdot
           {40h_A^4\over 81 f_\pi^2} {1\over(4\pi f_\pi)^2 }
             F_1(\eta,-\delta)   \nonumber \\
           & & \  \ \ \ \ \
               \times  \Big[M\rlap/q  - \bar{M}\eta
                        -\case{1}{20}(2m_\pi^2 - t -12\eta^2)\Big]
\;,\\[5pt]
   T^{ba}_{9l} &=&\Big(\delta^{ab} +\case{1}{6}[\tau^b,\tau^a]\Big)\cdot
           {100h_A^2\over 729 f_\pi^2} 
                        {\tilde{h}_A^2\over(4\pi f_\pi)^2 }
                  F_1(\eta-\delta, -\delta)     \nonumber \\
           & & \  \ \ \ \ \
               \times  \Big[M\rlap/q  - \bar{M}\eta
                        -\case{1}{2}(2m_\pi^2 - t -3\eta^2)\Big]
\;,\\[5pt]
   T^{ba}_{9m} &=&\Big(\delta^{ab} +\case{1}{2}[\tau^b,\tau^a]\Big)\cdot
           {3g_A^4\over 4 f_\pi^2} {1 \over(4\pi f_\pi)^2 }
            \Big( M\rlap/q  - \bar{M}\eta
                        +\thalf \eta^2\Big) F_1(\eta,0)
\;,\\[5pt]
   T^{ba}_{9n} &=&\Big(\delta^{ab} +\case{1}{2}[\tau^b,\tau^a]\Big)\cdot
           {4g_A^2\over 3 f_\pi^2} {h_A^2 \over(4\pi f_\pi)^2 }
            \Big( M\rlap/q  - \bar{M}\eta
                        +\thalf \eta^2\Big) F_1(\eta-\delta, -\delta)
\;,\\[5pt]
   T^{ba}_{9o} &=&-\case{1}{2}[\tau^b,\tau^a]\cdot
           {5\over 24 f_\pi^2} {1 \over(4\pi f_\pi)^2 }
                    m_\pi^2\eta
                      \ln{m_\pi^2 \over \mu^2}
\;,\\[5pt]
   T^{ba}_{9p} &=&-\case{1}{2}[\tau^b,\tau^a]\cdot
           {g_A^2\over 8 f_\pi^2} {1 \over(4\pi f_\pi)^2 }
                    m_\pi^2\eta
                      \bigg(2+ 3\ln{m_\pi^2 \over \mu^2}
                             \bigg) 
\;,\\[5pt]
   T^{ba}_{9q} &=&\case{1}{2}[\tau^b,\tau^a]\cdot
           {20 h_A^2\over 9 f_\pi^2} {\eta \over(4\pi f_\pi)^2 }
                    \bigg[ m_\pi^2 - 2 \delta^2 +3 \delta J(\delta)
                         + \case{3}{2}m_\pi^2\ln{m_\pi^2 \over \mu^2}
                             \bigg]
\;,\\[5pt]
   T^{ba}_{9r} &=&  T^{ba}_{9s}  = 
            - \delta^{ab} \cdot {\pi g_A^2 \over 3 f_\pi^2}  
                               {m_\pi^3  \over(4\pi f_\pi)^2 }
\;,\\[5pt]
   T^{ba}_{9t} &=&  T^{ba}_{9u}  = \delta^{ab} \cdot
            {16 h_A^2 \over 27 f_\pi^2} {1 \over(4\pi f_\pi)^2 }
              \bigg[ (\delta^2-m_\pi^2) J(\delta) + \thalf 
                  m_\pi^2\delta \ln{m_\pi^2 \over \mu^2} \bigg]
\;,\\[5pt]
   T^{ba}_{10a} &=&\Big(\delta^{ab} +\case{1}{2}[\tau^b,\tau^a]\Big)\cdot
           {2g_A\lambda_1m_\pi^2 \over M^2f_\pi^2\eta}
            \Big( M\rlap/q  - \bar{M}\eta\Big)
\;,\\[5pt]
   T^{ba}_{10b} &=& \Big(\delta^{ab} +\case{1}{2}[\tau^b,\tau^a]\Big)\cdot
           {g_A^2\over 3 f_\pi^2} {1 \over(4\pi f_\pi)^2 }
            \Big( M\rlap/q  - \bar{M}\eta
                        +\thalf \eta^2\Big)
                     {m_\pi^2 \over \eta}  \ln{m_\pi^2 \over \mu^2} 
\;,\\[5pt]
   T^{ba}_{10c} &=& -\Big(\delta^{ab} +\case{1}{2}[\tau^b,\tau^a]\Big)\cdot
           {g_A^4\over 6 f_\pi^2} {1 \over(4\pi f_\pi)^2 \eta}
            \Big( M\rlap/q  - \bar{M}\eta
                        +\thalf \eta^2\Big) F_3(\eta,0)
\;,\\[5pt]
   T^{ba}_{10d} &=& -\Big(\delta^{ab} +\case{1}{2}[\tau^b,\tau^a]\Big)\cdot
           {32g_A^2 \over 27 f_\pi^2} {h_A^2 \over(4\pi f_\pi)^2\eta }
            \Big( M\rlap/q  - \bar{M}\eta
                        +\thalf \eta^2\Big) F_3(\eta,-\delta)
\;,\\[5pt]
   T^{ba}_{10e} &=& -\Big(\delta^{ab} +\case{1}{2}[\tau^b,\tau^a]\Big)\cdot
           {32g_A^2 \over 27 f_\pi^2} {h_A^2 \over(4\pi f_\pi)^2\eta }
            \Big( M\rlap/q  - \bar{M}\eta
                        +\thalf \eta^2\Big)  F_3(\eta-\delta,0)
\;,\\[5pt]
   T^{ba}_{10f} &=& \Big(\delta^{ab} +\case{1}{2}[\tau^b,\tau^a]\Big)\cdot
           {200h_A^2 \over 243 f_\pi^2} 
              \, {g_A\tilde{h}_A \over(4\pi f_\pi)^2\eta }
            \Big( M\rlap/q  - \bar{M}\eta
                        +\thalf \eta^2\Big)  F_3(\eta-\delta,-\delta)
\;,\\[5pt]
   T^{ba}_{10g} &=&  -\Big(\delta^{ab} -\case{1}{4}[\tau^b,\tau^a]\Big)\cdot
              { 8h_A\lambda_6 \over 3f^2_\pi} {m_\pi^2 \over M^2}\,
           [\alpha_1(s,t)+\alpha_2(s,t) \rlap/{\mkern0mu}q]\Delta_R(s)
\;,\\[5pt]
   T^{ba}_{10h} &=&  -\Big(\delta^{ab} -\case{1}{4}[\tau^b,\tau^a]\Big)\cdot
            {4h_A^2\over 9 f_\pi^2}\, {1  \over(4\pi f_\pi)^2 }
              m_\pi^2 \ln{m_\pi^2 \over \mu^2}\,
        \Big[ \alpha_1(s,t)+ \alpha_2(s,t)\rlap/q\Big]
                    \Delta_R(s)
\;,\\[5pt]
   T^{ba}_{10i} &=& \Big(\delta^{ab} -\case{1}{4}[\tau^b,\tau^a]\Big)\cdot
       {8g_A^2\over 9 f_\pi^2}\, {h_A^2 \over(4\pi f_\pi)^2 }
                 \Big[ \alpha_1(s,t)+\alpha_2(s,t)\rlap/q\Big]
                                       \nonumber \\
           & & \  \ \ \ \ \
               \times 
                \bigg \{ F_3(\eta,0)  \Delta_R(s)
                  + {12\pi^2f_\pi^2 \over h_A^2\eta}\,
                     {M_\Delta
                       \Gamma_\Delta^2(s) 
                     \over
               [s - M_\Delta^2 - \Pi_\Delta(\eta)]^2 +
                    M_\Delta^2\Gamma_\Delta^2(s) } \bigg \}
\;,           \label{eq:10g} \\[5pt]
   T^{ba}_{10j} &=&\Big(\delta^{ab} -\case{1}{4}[\tau^b,\tau^a]\Big)\cdot
            {8h_A^4\over 81 f_\pi^2}\, {1 \over(4\pi f_\pi)^2 }
                 \Big[ \alpha_1(s,t)+\alpha_2(s,t)\rlap/q\Big]
                                   \nonumber \\
           & & \  \ \ \ \  \ \ \
               \times  \bigg \{  F_3(\eta,-\delta)  \Delta_R(s)
                  + {12\pi^2 f_\pi^2\over h_A^2(\eta+\delta)}\,
                     {  M_\Delta\Gamma_\Delta^2(s) 
                     \over
               [s - M_\Delta^2 - \Pi_\Delta(\eta)]^2 +
                    M_\Delta^2\Gamma_\Delta^2(s) } \bigg \}
\;,     \label{eq:10h}    \\[5pt]
   T^{ba}_{10k} &=& -\Big(\delta^{ab} -\case{1}{4}[\tau^b,\tau^a]\Big)\cdot
             {50h_A^2\over 81 f_\pi^2}\,
                    {g_A \tilde{h}_A \over(4\pi f_\pi)^2 }
               \Big[ \alpha_1(s,t)+ \alpha_2(s,t)\rlap/q\Big]
               F_3(\eta-\delta,0) \Delta_R(s)
\;,\\[5pt]
   T^{ba}_{10l} &=&\Big(\delta^{ab} -\case{1}{4}[\tau^b,\tau^a]\Big)\cdot
                {200\over 729}
                  {h_A^2\over f_\pi^2}\,
                    {\tilde{h}_A^2 \over(4\pi f_\pi)^2 }
          \Big[ \alpha_1(s,t)+\alpha_2(s,t)\rlap/q\Big]
           F_3(\eta-\delta,-\delta)  \Delta_R(s)\;.
\end{eqnarray}
Notice Eqs.~(A33) -- (A36) for diagrams 10(g)--10(l) have been written in 
terms of $\alpha_1,\ \alpha_2$ of Eq.~(\ref{alph}) rather than truncating 
these expressions to \ord{2}. This is necessary to ensure that the resonance 
only contributes to the $P_{33}$ channel and does not ``contaminate"
the other channels. To the above contributions should be added the 
results for the cross diagrams where applicable. They are
obtained from the listed expressions by the replacement $s\rightarrow u$,
$\eta \rightarrow \bar{\eta} \equiv  (u-M^2)/(2\bar{M})$,
and the interchanges $a\leftrightarrow b$ and $q\leftrightarrow-q'$.

\end{document}